\documentclass[prx,twocolumn,showpacs,superscriptaddress]{revtex4-1}
\usepackage{bm,graphicx,amsmath,amssymb,color,placeins,tikz,pgflibraryarrows,braket,colortbl,xcolor,mathtools}
\usepackage[caption=false]{subfig}
\usetikzlibrary{decorations.pathreplacing}
\def\D{\partial}
\def\DD{\Delta}

\def\<{\langle}
\def\>{\rangle}
\def\ve{\varepsilon}

\begin{document}

\title{Quantum Confinement of Electron-Phonon Coupling in Graphene Quantum Dots}

\author{Marios Zacharias} 
\email{marios.zacharias@cut.ac.cy}
\affiliation{Research Unit for Nanostructured Materials Systems,
  Cyprus University of Technology, P.O. Box 50329, 3603 Limassol,
  Cyprus}
\affiliation{Department of Mechanical and Materials Science
  Engineering, Cyprus University of Technology, P.O. Box 50329, 3603
  Limassol, Cyprus}

\author{Pantelis C. Kelires}
\affiliation{Research Unit for Nanostructured Materials Systems,
  Cyprus University of Technology, P.O. Box 50329, 3603 Limassol,
  Cyprus}
\affiliation{Department of Mechanical and Materials Science
  Engineering, Cyprus University of Technology, P.O. Box 50329, 3603
  Limassol, Cyprus}

\date{\today}

\begin{abstract}
On the basis of first-principles calculations and the special displacement method, 
we demonstrate the quantum confinement scaling law of the phonon-induced gap 
renormalization of graphene quantum dots (GQDs). We employ zigzag-edged GQDs with hydrogen passivation 
and embedded in hexagonal boron nitride. Our calculations for GQDs in the sub-10 nm region 
reveal strong quantum confinement of the zero-point renormalization ranging from 20 to 250 meV. 
To obtain these values we introduce a correction to the Allen–Heine theory of temperature-dependent 
energy levels that arises from the phonon-induced splitting of 2-fold degenerate edge states. 
This correction amounts to more than 50\% of the gap renormalization. We also present 
momentum-resolved spectral functions of GQDs, which are not reported in previous contributions. 
Our results lay the foundation to systematically engineer temperature-dependent electronic structures 
of GQDs for applications in solar cells, electronic transport, and quantum computing devices.

\end{abstract}

\maketitle

Owing to their quantum-confined band gap, graphene quantum dots (GQDs) 
exhibit intriguing photoluminescence and photocatalytic properties~\cite{Zhang2012} 
with potential applications in bioimaging~\cite{Chung2019,Younis2020}, photovoltaics~\cite{Gupta2011,Gao2014}, 
light-emitting diodes~\cite{Gupta2011,Kim2020}, and water splitting~\cite{Zeng2018,Yan2018}.
The highly tunable gap of GQDs also renders them as attractive nanomaterials for 
exploitation in the growing fields of nonlinear optics~\cite{Oluwole2018,Meng2018,Kuo2020} 
and quantum computing~\cite{Trauzettel2007,Eich_2018}. 
Similarly to graphene nanoribbons~\cite{Louie2006}, the gap ($\ve_g$) of pristine GQDs varies inversly proportional 
with the dot size ($L$), i.e. $\ve_g \propto 1/L$~\cite{Zhang_1998}. This behavior is well established. 
It has been validated by experimental measurements~\cite{Ritter_2009,Magda2014,CortsdelRo2020}
and the Dirac Fermion model~\cite{Zhang_2008}, as well as by tight-binding and 
first-principles calculations~\cite{Saleem_2019}. 
Apart from quantum confinement, the gap of nanometer-sized GQDs depends on the type of the edge, 
either zigzag, armchair, or hybrid, and the structure/shape of the 
dot~\cite{Kyoko1996,Ritter_2009,CortsdelRo2020,Yan2010,Pan2010,Wimmer2010,Ozfidan2015,Saleem_2019}. 
Combining and doping GQDs with functional groups and heteroatoms also 
provide fascinating routes for band structure engineering and, hence, harvesting
novel photophysics for technological 
applications~\cite{Li2011,Yan2011,Qian2013,Wang2016,Wang2016_2,Qian2016,Yan2018,Bayoumy2019,Kadian_2019}.

Along with the rapid progress in understanding the properties of GQDs, attention has also been 
devoted to the development of GQDs embedded in hexagonal boron nitride 
(GQDs/h-BN)~\cite{Ci2010,Peng2013,Kang2013,Liu2013,Liu2014,Kim2015,Ghahari2017,Chen_2019}. 
The similar bond length of monolayer h-BN with graphene facilitates the synthesis of stable 
in-plane heterostructures with controlled GQDs domains. 
The advantage of fabricating GQD/h-BN is that undesired adsorption on the carbon 
dangling bonds and graphene edge reconstruction are eliminated~\cite{Pekka2008,Girit2009,Kim2013}. 
This allows to stabilize graphene domains, opening the way to investigate the intrinsic properties of GQDs.
First-principles studies~\cite{Bhowmick2011,Li2011_b,Zhao2013} have demonstrated 
that GQDs/h-BN also follow a band gap scaling law of $\ve_g \propto 1/L$, preserving 
the quantum confined states of graphene Dirac-like electrons.

Despite the advance in electronic structure calculations of GQDs, either free-standing (FS) 
or embedded, focus has been placed on their ground, or excited, state properties at $0$~K, 
where nuclei are treated as classical particles. In the last two decades, significant 
developments have been made in incorporating successfully the effect of quantum nuclear dynamics 
in ab-initio calculations~\cite{FG_review}. Those include methods within the framework of 
perturbative~\cite{Marini2008,Giustino_2010,Ponce_2014,Antonius_2014,Ponce_2015,Lihm_2020,Gonze_2020} 
and non-perturbative approaches~\cite{Patrick_2013,Patrick_2014,Bartomeu_2014,Zacharias_2015,
Zacharias_2016,Bartomeu_2016,Karsai_2018,Zacharias_2020} 
for the treatment of electron-phonon coupling; as well as of molecular 
dynamics~\cite{Della_Sala_2004,Ramirez_2006,Ramirez_2008,Zacharias_2020_b,Galli_2021} and
quantum Monte Carlo~\cite{Gorelov2020,Hunt_2020} techniques for the treatment of vibronic interactions.

\begin{figure}[t!]
\includegraphics[width=0.50\textwidth]{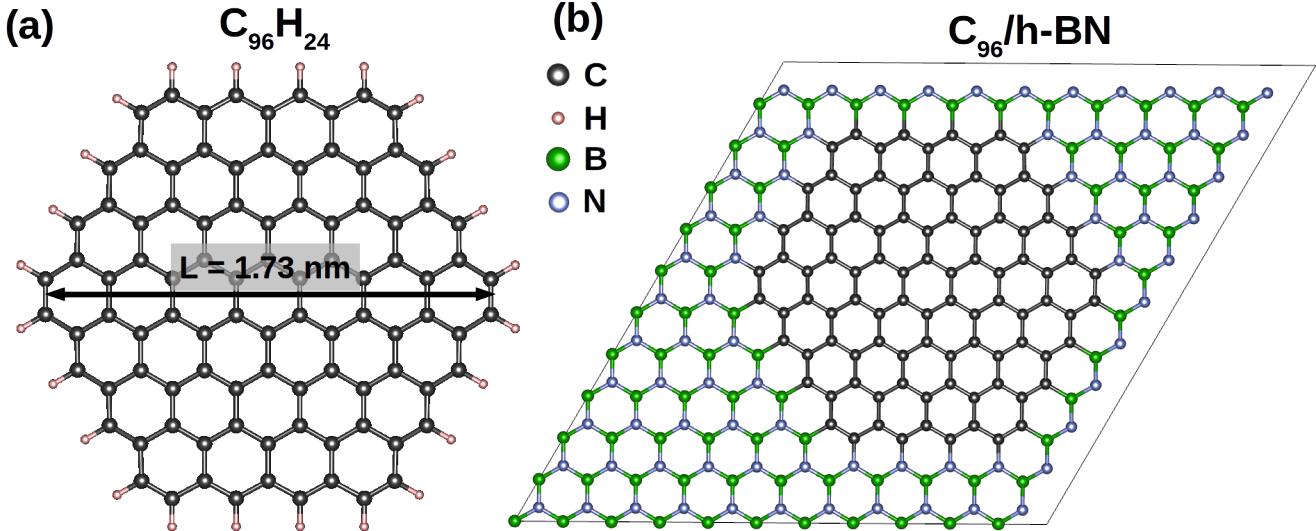}
\caption{\label{fig1}(a), (b) Simulation cells of zigzag-edged hexagonal GQD passivated with 
hydrogen (H) atoms and embedded in a rhombohedral h-BN superlattice. 
Each GQD contains 96 carbon (C) atoms having a lateral dimension $L = 1.73$ nm.}
\end{figure}

In this Letter, we employ the special displacement method (SDM)~\cite{Zacharias_2016,Zacharias_2020}, 
an efficient non-perturbative approach, in conjunction with degenerate perturbation theory 
to investigate quantum confinement effects on the phonon-induced gap renormalization
of zigzag free-standing (Figure~\ref{fig1}a) and embedded in h-BN (Figure~\ref{fig1}b) GQDs. 
The band extrema of zigzag FS-GQDs are characterized by a 2-fold 
degeneracy~\cite{Saleem_2019}, arising from the rotational symmetry 
of the system~\cite{Ozfidan2015}. The degeneracy is preserved upon embedding in h-BN. 
However, perturbations due to thermal vibrations lift the degeneracy and 
the gap renormalization needs to be determined by means of degenerate perturbation theory. 
Within the harmonic and adiabatic approximations, the renormalization of a 2-fold degenerate energy level 
at temperature $T$, and up to second order in normal coordinates is given by: 
\begin{eqnarray}\label{eq.ephcc_AH}
          \DD\ve_{c_1,T}^{\pm} =  &\pm& \bigg| \sum_\nu (g_{c_1c_1 \nu} - g_{c_2c_2 \nu}) \frac{\sigma_{\nu,T}}{\sqrt{2 \pi}}\bigg|
\\ \nonumber &+&  
{\sum_{\nu \beta}}^\prime \bigg[ \frac{|g_{c_1 \beta\nu}|^2} {\ve_{c_1}-\ve_\beta} + h_{c_1 \nu}\bigg] \sigma^2_{\nu,T}.
\end{eqnarray}
Here $c_1$ and $c_2$ are the indices of the 2-fold degenerate states defined by energies $\ve_{c_1}$, $\ve_{c_2}$
and wavefunctions $\psi_{c_1}$, $\psi_{c_2}$. Both summations run over the normal modes $\nu$, while 
the primed summation is also taken over the energies except for $\ve_\beta = \ve_{c_1}, \ve_{c_2}$.
The electron-phonon matrix elements are given by $g_{c_1 \beta\nu} = \< \psi_{c_1} | \D V_{\rm KS}/\D x_\nu | \psi_\beta\> $
and $h_{c_1 \nu} = 1/2 \, \< \psi_{c_1} | \D^2 V_{\rm KS}/\partial x^2_\nu | \psi_{c_1}\>$, where
 $V_{\rm KS}$ represents the Kohn-Sham potential and $x_\nu$ the normal coordinates.
The temperature dependence of the energy levels comes from 
$\sigma^2_{\nu,T} = l^2_\nu (2 n_{\nu,T} + 1)$, where $l_\nu$ is the zero-point 
amplitude~\cite{Zacharias_2020} and $n_{\nu,T} = [\exp(\hbar \omega_\nu/k_{\rm B} T) - 1]^{-1}$ 
is the Bose-Einstein occupation of the phonon with frequency $\omega_\nu$. 
The second line of Eq.~\eqref{eq.ephcc_AH} represents the standard energy renormalization 
described within the framework of non-degenerate Allen-Heine (AH) theory~\cite{Allen_1976,Allen_Cardona_1981}. 
The first and second terms inside the square brackets are recognized as the Fan-Migdal 
and Debye-Waller contributions. The term in the first line of Eq.~\eqref{eq.ephcc_AH} is 
the {\it correction} to the AH theory for finite systems exhibiting 2-fold degenerate band extrema, 
and the $\pm$ signs define the upper and lower states in energy arising from nuclei 
thermal vibrations. The derivation of Eq.~\eqref{eq.ephcc_AH} is available in the Supporting Information.
We remark that in the harmonic approximation, the degeneracy splitting reduces 
to zero at the thermodynamic limit~\cite{Zacharias_2020} and therefore this correction can be safely 
ignored for extended systems.

Now we recall~\cite{Bhowmick2011} that for GQDs, $\psi_{c_1}$ and $V_{\rm KS}$ are proportional to 
$1/L$ and $L$, respectively, to readily derive the scaling laws of the electron-phonon matrix elements: 
$g_{c_1 \beta \nu} \propto 1/L$ and $h_{c_1 \nu} \propto 1/L$. Furthermore, we make the assumption 
that the Fan-Migdal contribution in Eq.~\eqref{eq.ephcc_AH} is dominated by the terms with 
$\ve_g = \ve_{c_1}-\ve_{v_1}$ in the denominator. This leads to the following quantum 
confinement law: 
\begin{eqnarray}\label{eq.ephcc_AH_law_g_h}
      \DD \ve^{\pm}_{c_1,T } \propto \frac{1}{L}. 
\end{eqnarray}

To compute the energy level renormalization of GQDs we employ SDM, developed by Zacharias and 
Giustino (ZG)~\cite{Zacharias_2016,Zacharias_2020}. In this approach, a single
distorted configuration of the system that captures the effect of electron-phonon 
coupling is generated using ZG displacements. For the purposes of this work, we combine 
ZG displacements together with their antithetic set~\cite{Zacharias_2016} to evaluate: 
\begin{eqnarray}\label{eq.ephcc_SDM}
      \DD \ve^{\pm}_{c_1,T } = &\pm& \bigg| \sum_\nu 
     \bigg( \frac{\partial \ve_{c_1}}{\partial x_\nu} - \frac{\partial \ve_{c_2}}{\partial x_\nu} \bigg)
     \frac{\sigma_{\nu,T}}{\sqrt{2 \pi}} \bigg| \\
       &+& \frac{1}{2}\sum_\nu \frac{\partial^2 \ve_{c_1}}{\partial x^2_\nu} \sigma^2_{\nu,T}
        + \mathcal{O}(\sigma_{\nu,T}^4),
\nonumber
\end{eqnarray}
where $\mathcal{O}(\sigma_{\nu,T}^4)$ represents terms of fourth order 
in normal coordinates and higher, contributing to the phonon-induced renormalization. 
Terms up to quadratic order are equivalent to those in Eq.~\eqref{eq.ephcc_AH}. 
The electron-phonon matrix elements 
are not computed explicitly as in perturbative methods~\cite{Ponce_2016_EPW}, 
but inherently included via the derivatives 
$\partial \ve_{c_1} / \partial x_\nu$ and $\partial^2 \ve_{c_1} / \partial x^2_\nu$. 
More computational details are available in the Supporting Information.

\begin{figure*}[hbt!]
\includegraphics[width=0.93\textwidth]{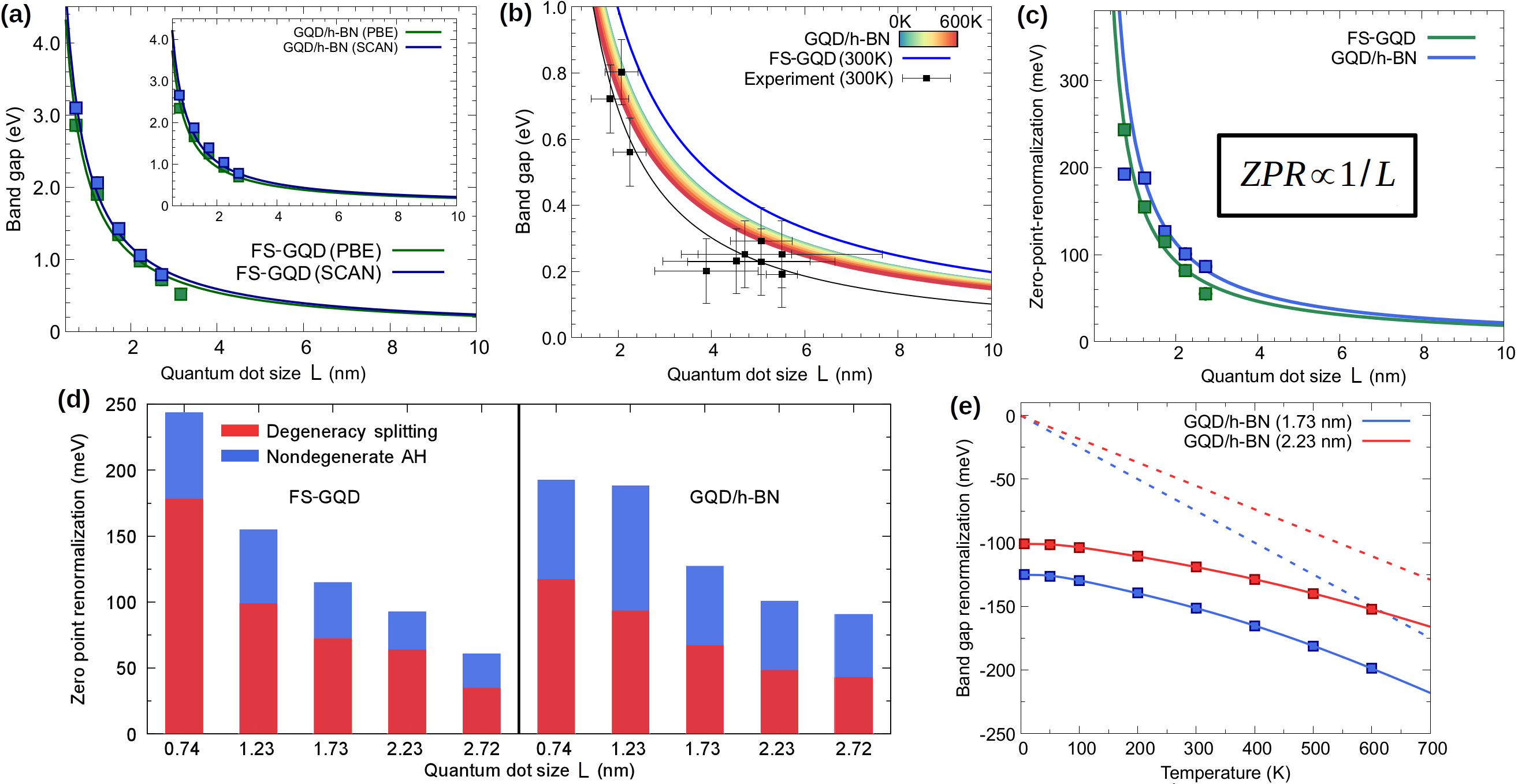}
\caption{\label{fig2}
  (a) Gap of FS-GQDs versus $L$ calculated with the nuclei at static 
  equilibrium using DFT-PBE (blue) and DFT-SCAN (green). In the inset, 
  data for GQD/h-BN is shown. The solid curves are fits of the form $a/L$ to the calculated gaps.
  (b) Fits of the form $a/L$ to the calculated gaps of GQDs at finite temperatures. 
  Blue curve represents the fit for FS-GQDs at 300~K and the color map 
  represents the fits for GQDs/h-BN in the range 0~-- 600~K.
  Experimental data from ref.~\cite{Ritter_2009}  and the fit of the form 1.57~eV~nm/$L^{1.19}$ are shown as black.
  (c) Zero-point renormalization (ZPR) as a function of $L$ calculated for FS-GQDs (green) and 
  GQDs/h-BN (blue). The curves represent fits to $c/L$. 
  Values of $a$ and $c$ are given in the main text. 
  (d) Decomposition of the zero-point renormalization into
  the non-degenerate AH (blue) and degeneracy splitting (red) contributions. 
  Left and right panels show data of five different sized FS-GQDs and GQDs/h-BN.  
  (e) Temperature-dependent band gap renormalization of GQD/h-BN with size 1.73 nm (red) and 2.23 nm (blue). 
  Solid curves are fits to double Bose-Einstein oscillators [Eq.~\eqref{eq.double_BE}] and dashed lines 
  are the high-temperature asymptotes. 
  All calculations in (b)-(e) are performed using DFT-PBE. 
}
\end{figure*}

Figure \ref{fig2}a shows the bare gap energy, $\ve_g$, as a function of the GQD size, $L$. 
Calculations were performed using density-functional theory (DFT) in the PBE 
generalized gradient approximation~\cite{GGA_Pedrew_1996}  
and SCAN meta-generalized gradient approximation~\cite{SCAN_2015} as implemented in 
Quantum Espresso~\cite{QE,QE_2} and LibXC library~\cite{LIBXC_2018}. 
For our calculations we employed the fully relaxed zigzag-edged structures of size up to $L=3.16$~nm. 
Details for the structures and calculations are available in the Supporting Information. 
The solid curves represent fits to the data of the form $a/L$, as expected for Dirac fermions. 
Our fits to FS-GQDs gaps give $a^{\rm PBE}_{\rm FS} = 2.18$~eV~nm and $a^{\rm SCAN}_{\rm FS} = 2.35$~eV~nm
with an asymptotic error at 2.5\%. 
For embedded GQDs, we obtain $a^{\rm PBE}_{\rm h-BN} = 1.87$~eV~nm and $a^{\rm SCAN}_{\rm h-BN} = 2.11$~eV~nm
with an asymptotic error at 4.5\%.
To make contact with the literature, we also fit our data for GQDs/h-BN to the power law $b/\sqrt{N}$, where $N$ 
is the number of carbon atoms in each structure. Our analysis yields $b^{\rm PBE}_{\rm h-BN} = 11.6$~eV 
and $b^{\rm SCAN}_{\rm h-BN} = 13.2$~eV in agreement with the value of $15.0$~eV obtained 
from tight-binding data~\cite{Zhao2013}. Our PBE and SCAN 
band gaps compare well, with discrepancies being less than 100~meV for $L>1.7$~nm.
Furthermore, the gaps of GQDs/h-BN are systematically lower than those
calculated for FS-GQDs. This result demonstrates that, apart from quantum confinement,
the gap of GQDs is affected by embedding, and more particularly, by the 
participation of B and N 2p orbitals in the formation of the band edges~\cite{Li2011_b}
(see Supporting Information, Figure~S2).

In Figure~\ref{fig2}b we present fits of the form $a(T)/L$ to the band gaps at finite 
temperatures calculated using ZG displacements for free-standing and embedded GQDs 
and we compare with experiment~\cite{Ritter_2009}. Temperature-dependent band gaps are 
obtained from: 
\begin{eqnarray}\label{eq.ephcc_BG}
    \ve_{g,T} = \ve_g + \frac{\DD \ve^{-}_{c_1,T } + \DD \ve^{-}_{c_2,T } - \DD \ve^{+}_{v_1,T } - \DD \ve^{+}_{v_2,T } }{2}, 
\end{eqnarray}
where $c_1$,$c_2$ and $v_1$,$v_2$ indicate the degenerate states of the 
conduction band minimum (CBM) and valence band maximum (VBM). 
Our analysis yields $a^{\rm PBE}_{\rm FS} (300~{\rm K}) = 1.98 \pm 0.05$~eV~nm in line 
with the experimental value $a^{\rm expt} (300~{\rm K}) = 1.57\pm 0.21$~eV~nm. 
This comparison demonstrates that accounting for electron-phonon coupling in our calculations 
improves the agreement with experiment. 
The remaining discrepancies can be attributed to (i) non-adiabatic effects, not included in our 
calculations (see below) and (ii) shortcomings in identifying
well-defined edge structures and shapes of GQDs in sub-10~nm~\cite{Ritter_2009,Magda2014}. 
Modern techniques based on scanning tunnelling microscopy and hydrogen patterning~\cite{CortsdelRo2020} 
can be proved useful to improve comparison. 
The colour map in Figure~\ref{fig2}b shows $a(T)/L$ obtained for GQDs/h-BN 
at $T= 0$ -- $600$~K. The parameters $a^{\rm PBE}_{\rm h-BN} (T)$ lie in 
the range $1.69$ -- $1.49$~eV~nm (fitting error at 4\%), 
demonstrating that the gap of GQDs decreases with temperature as for
conventional crystalline semiconductors~\cite{Ponce_2015,Zacharias_2020}.
Importantly, our calculated and extrapolated band gaps are consistent with measurements 
on GQDs/h-BN, showing no photoluminescence in the visible spectrum~\cite{Chen_2019}.

\begin{figure*}[htb!]

\includegraphics[width=0.99\textwidth]{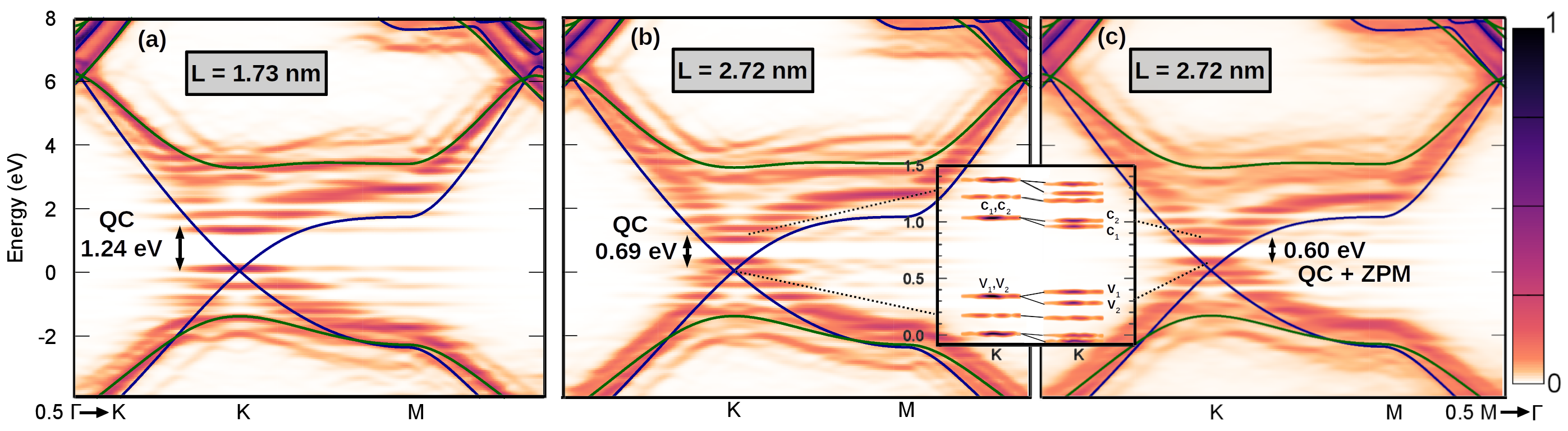}
\caption{\label{fig3} Momentum-resolved spectral functions calculated for GQDs/h-BN of:
(a) $L=1.73$~nm and with atoms at static equilibrium, (b) $L=2.72$~nm and 
with atoms at static equilibrium, and (c) $L=2.72$~nm and with atoms at 
their ZG configuration generated for $T=0$~K. The band gap is indicated on each plot. 
QC and ZPM stand for quantum confinement and zero-point motion. Blue and green 
lines represent the band structures of graphene and monolayer h-BN, respectively. 
The inset shows the energy levels around the band edges of (b) and (c) at the 
$K$ high-symmetry point.} 
\end{figure*}

Figure~\ref{fig2}c shows the zero-point renormalization of the band gap of free-standing (green) 
and embedded (blue) GQDs as a function of $L$. Fitting our data to $c/L$ yields 
$c^{\rm PBE}_{\rm FS} = 183$~meV~nm and $c^{\rm PBE}_{\rm h-BN} = 227$~meV~nm 
with an asymptotic error for each at around 2.0\%.
This result together with the data in Figure~\ref{fig2}b 
verify the scaling law derived in Eq.~\eqref{eq.ephcc_AH_law_g_h} and justify
that the electron-phonon coupling in GQDs is subject to quantum confinement.
It is also evident that embedding in h-BN enhances the effect of electron-phonon coupling, 
as a result of the localized $\pi$ states at the interface between 
the dot and the h-BN matrix, which modify the wavefunctions of the edge states. 
This behavior is at variance with calculations for silicon nanocrystals where embedding in 
amorphous silica strongly suppresses electron-phonon coupling and zero-point renormalization~\cite{Zacharias_2020_3}.
We note that the zero-point renormalization calculated for the GQD/h-BN 
of $L=0.74$~nm deviates by $100$~meV from the quantum confinement law $c/L$ 
and, therefore, it was excluded from the fitting procedure. We attribute 
this deviation to the fact that an enhanced portion of the VBM and CBM charge densities 
is localized at the h-BN network (see Supporting Information, Figure~S2). 

We now test the role of non-adiabatic effects~\cite{Ponce_2015,Gonze_2020} in
the zero-point renormalization by modifying the amplitude of the 
ZG displacements via $\sigma^{\rm na}_{\nu,T} = \sqrt{\ve_g / (\ve_g - \hbar \omega_\nu)} \sigma_{\nu,T}$; 
we justify this choice in the Supporting Information.
These effects are known to be significant in materials whose energy gap is 
of the order of phonon energies. Our calculation for the FS-GQD of 
$L=2.23$~nm yields a zero-point renormalization of 112~meV, 
amounting to a further decrease of the gap by 13~meV. We point out that non-adiabaticity 
is expected to dominate phonon-induced renormalization for $L > 6$~nm, leading to deviations 
from the scalling law $\ve_{g,T} \propto 1/L$. This conclusion might explain
measurements~\cite{Ritter_2009} revealing a zero gap for GQDs of $L > 6$~nm. 
To clarify this aspect further calculations within the non-adiabatic AH theory are essential.

In Figure~\ref{fig2}d we present the decomposition of the zero-point renormalization 
into contributions arising from degeneracy splitting (red) and non-degenerate AH theory (blue). 
Those stem from the first and second lines of Eq.~\eqref{eq.ephcc_SDM}, respectively. For 
FS-GQDs, degeneracy splitting dominates zero-point renormalization amounting
up to 73\%. The same contribution calculated for embedded GQDs is around 50\%.
We emphasize that both degeneracy splitting and non-degenerate AH renormalizations 
decrease with the GQD's size, scaling as $1/L$ 
(apart from the AH term obtained for GQD/h-BN with $L=0.74$~nm).  
We note that in ref.~\cite{Galli_2021}, an energy splitting
of $\sim$ 200~meV has been reported for pentamantane C$_{26}$H$_{32}$ within the
frozen-phonon harmonic approximation. Pentamantane exhibits a 3-fold degenerate 
highest occupied level and determining its gap renormalization requires 
degenerate perturbation theory. In the Supporting Information, we generalize our 
formalism and suggest the treatment of phonon-induced renormalization of an 
$n$-fold degenerate energy level. 

Figure~\ref{fig2}e shows the band gap renormalization as a function of temperature
evaluated for GQDs/h-BN with $L=1.73$~nm (blue) and $2.23$~nm (red). 
Solid curves represent double Bose-Einstein oscillators of the form: 
\begin{eqnarray}\label{eq.double_BE}
\Delta\ve_{g,T} = -(\alpha_1+\alpha_2)-\frac{2\alpha_1}{e^{\epsilon_1/k_{\rm B}T}-1}
                                      -\frac{2\alpha_2}{e^{\epsilon_2/k_{\rm B}T}-1}. 
\end{eqnarray}
Here $\alpha_1$, $\alpha_2$ represent the contributions to the zero-point renormalization
due to effective phonons with energies $\epsilon_1$, $\epsilon_2$. 
Our fits yield ($\alpha_1=115$, $\epsilon_1 = 146$), ($\alpha_2 = 10$, $\epsilon_2 = 15$)~meV and
($\alpha_1=93$, $\epsilon_1 = 159$), ($\alpha_2 = 8$, $\epsilon_2 = 17$)~meV
for the structures with $L=1.73$~nm and $2.23$~nm, respectively. 
These values demonstrate that the renormalization arises mainly from the
coupling of electrons to in-plane optical phonons.
The high temperature asymptotes 
are described by 0.25~$T$~meV/K (blue) and 0.18~$T$~meV/K (red) and 
can be interpreted as the classical limit of $\Delta\ve_{g,T}$.

In Figure~\ref{fig3} we report the momentum-resolved spectral function 
of GQDs/h-BN obtained using the band structure unfolding technique~\cite{Popescu_2012,Medeiros_2014} 
as implemented in the {\tt EPW} software package~\cite{Ponce_2016_EPW,Zacharias_2020}. 
For our calculations we employed the static equilibrium structures of size $1.73$~nm (Figure~\ref{fig3}a) 
and $2.72$~nm (Figure~\ref{fig3}b), as well as the ZG configuration of size $2.72$~nm 
for $T=0$~K (Figure~\ref{fig3}c). At variance with the electronic band structures of pristine graphene 
(blue lines) and monolayer h-BN (green lines), the spectral functions of GQDs/h-BN exhibit 
nondispersive band edges. This justifies our key assumption leading to Eq.~\eqref{eq.ephcc_AH_law_g_h}, 
i.e. to replace the energy denominators of the Fan-Migdal term in Eq.~\eqref{eq.ephcc_AH} with $\ve_g$.
Our calculations for the static equilibrium structures indicate that the VBM increases by 
0.38~eV while the CBM decreases by 0.17~eV when $L$ changes from 
$1.73$~nm to $2.72$~nm. Furthermore, the density of states is enhanced around the band edges, showing 
that more features of graphene's band structure are recovered as $L$ increases. 
Inclusion of quantum zero-point motion in our calculations
causes narrowing of the gap and splitting of the degenerate energy levels, as shown
in the inset of Figure~\ref{fig3}. In particular, ZG displacements at 0~K cause a splitting 
of 98~meV and 49~meV of the VBM and CBM. We emphasize that a correct evaluation 
of the splitting correction to AH theory 
requires calculations using antithetic ZG displacements (see Supporting Information).
We also stress that this lifting of energy degeneracy might help interpreting 
the applications of GQDs in sub-10~nm as spin qubits~\cite{Freitag2016,Vandersypen2017}. 

The spectral functions in Figure~\ref{fig3} can serve as a starting point and the basis 
to methodically assess the effects of doping, adsorption, quantum confinement, and 
electron-phonon coupling on the electronic structure of GQDs. For example, GQDs/h-BN of $L<3$~nm
feature a flattening of the Dirac cone which limits the electron group velocity and therefore 
their transport properties. Adjusting the size and the shape of the dot, as well as doping 
with heteroatoms can lead to a continuum of dispersive states at the band edges with a desirable 
band gap. Hence, improved transport properties of GQDs can be achieved, making them favorable 
electrode materials for applications, e.g, in lithium ion batteries. Optimization of the
electronic structure is also critical to accelerate research on the use of GQDs in 
photovoltaics~\cite{Gupta2011}, as well as in quantum-logic devices with long coherence times~\cite{Trauzettel2007}.

In conclusion, we proposed a new approach for the calculation of temperature-dependent 
energy gaps of quantum dots that relies on the special displacement method and degenerate 
perturbation theory. We demonstrated a new quantum confinement law for the gap renormalization 
of GQDs, either free-standing or embedded in h-BN. We also introduced a correction to 
the non-degenerate AH theory arising from degeneracy splitting of edge states. This finding 
can be extended to explore electron-phonon coupling effects in other technologically 
important nanostructures, like halide perovskites and conventional CdSe and PbSe 
quantum dots~\cite{Quarti2020}, exhibiting degeneracies.  
Lastly, we revealed full momentum-resolved spectral functions of GQDs/h-BN, including 
> the effect of quantum zero point motion.
Our results offer a new route for systematic band structure engineering to 
improve optical, transport, and quantum computing properties of GQDs.  
Furthermore, our work can be upgraded to investigate exciton-phonon
spectra of nanostructures via the combination of SDM with the Bethe-Salpeter 
equation~\cite{Huang2021}. 

Electronic structure calculations performed in this study
are available on the NOMAD repository~\cite{nomad_doi}.

\section{Acknowledgments}
This work was supported by the Research Unit of Nanostructured Materials Systems (RUNMS)
and the program META$\Delta$I$\Delta$AKT$\Omega$P of the Cyprus University of Technology.
The results of this research have been achieved using the DECI resource Saniyer at 
UHeM [http://en.uhem.itu.edu.tr] with support from the PRACE aisbl and computational time 
provided by the HPC Facility of the Cyprus Institute [http://hpcf.cyi.ac.cy]. 
\\ \ \\
{\bf Supporting Information Available:} General computational details, derivation of the correction to the 
Allen-Heine theory in the case of 2-fold degenerate band extrema, charge densities of the VBM and 
CBM of GQDs/h-BN, evaluation of Eq.~\eqref{eq.ephcc_SDM} employing antithetic pairs of ZG displacements, 
and inclusion of non-adiabatic effects using the special displacement method.

\bibliography{references}{} 

\begin{thebibliography}{86}%
\makeatletter
\providecommand \@ifxundefined [1]{%
 \@ifx{#1\undefined}
}%
\providecommand \@ifnum [1]{%
 \ifnum #1\expandafter \@firstoftwo
 \else \expandafter \@secondoftwo
 \fi
}%
\providecommand \@ifx [1]{%
 \ifx #1\expandafter \@firstoftwo
 \else \expandafter \@secondoftwo
 \fi
}%
\providecommand \natexlab [1]{#1}%
\providecommand \enquote  [1]{``#1''}%
\providecommand \bibnamefont  [1]{#1}%
\providecommand \bibfnamefont [1]{#1}%
\providecommand \citenamefont [1]{#1}%
\providecommand \href@noop [0]{\@secondoftwo}%
\providecommand \href [0]{\begingroup \@sanitize@url \@href}%
\providecommand \@href[1]{\@@startlink{#1}\@@href}%
\providecommand \@@href[1]{\endgroup#1\@@endlink}%
\providecommand \@sanitize@url [0]{\catcode `\\12\catcode `\$12\catcode
  `\&12\catcode `\#12\catcode `\^12\catcode `\_12\catcode `\%12\relax}%
\providecommand \@@startlink[1]{}%
\providecommand \@@endlink[0]{}%
\providecommand \url  [0]{\begingroup\@sanitize@url \@url }%
\providecommand \@url [1]{\endgroup\@href {#1}{\urlprefix }}%
\providecommand \urlprefix  [0]{URL }%
\providecommand \Eprint [0]{\href }%
\providecommand \doibase [0]{http://dx.doi.org/}%
\providecommand \selectlanguage [0]{\@gobble}%
\providecommand \bibinfo  [0]{\@secondoftwo}%
\providecommand \bibfield  [0]{\@secondoftwo}%
\providecommand \translation [1]{[#1]}%
\providecommand \BibitemOpen [0]{}%
\providecommand \bibitemStop [0]{}%
\providecommand \bibitemNoStop [0]{.\EOS\space}%
\providecommand \EOS [0]{\spacefactor3000\relax}%
\providecommand \BibitemShut  [1]{\csname bibitem#1\endcsname}%
\let\auto@bib@innerbib\@empty
\bibitem [{\citenamefont {Zhang}\ \emph {et~al.}(2012)\citenamefont {Zhang},
  \citenamefont {Zhang}, \citenamefont {Chen},\ and\ \citenamefont
  {Qu}}]{Zhang2012}%
  \BibitemOpen
  \bibfield  {author} {\bibinfo {author} {\bibfnamefont {Z.}~\bibnamefont
  {Zhang}}, \bibinfo {author} {\bibfnamefont {J.}~\bibnamefont {Zhang}},
  \bibinfo {author} {\bibfnamefont {N.}~\bibnamefont {Chen}}, \ and\ \bibinfo
  {author} {\bibfnamefont {L.}~\bibnamefont {Qu}},\ }\href {\doibase
  10.1039/c2ee22982j} {\bibfield  {journal} {\bibinfo  {journal} {Energy
  Environ. Sci.}\ }\textbf {\bibinfo {volume} {5}},\ \bibinfo {pages} {8869}
  (\bibinfo {year} {2012})}\BibitemShut {NoStop}%
\bibitem [{\citenamefont {Chung}\ \emph {et~al.}(2019)\citenamefont {Chung},
  \citenamefont {Revia},\ and\ \citenamefont {Zhang}}]{Chung2019}%
  \BibitemOpen
  \bibfield  {author} {\bibinfo {author} {\bibfnamefont {S.}~\bibnamefont
  {Chung}}, \bibinfo {author} {\bibfnamefont {R.~A.}\ \bibnamefont {Revia}}, \
  and\ \bibinfo {author} {\bibfnamefont {M.}~\bibnamefont {Zhang}},\ }\href
  {\doibase 10.1002/adma.201904362} {\bibfield  {journal} {\bibinfo  {journal}
  {Adv. Mater.}\ }\textbf {\bibinfo {volume} {33}},\ \bibinfo {pages} {1904362}
  (\bibinfo {year} {2019})}\BibitemShut {NoStop}%
\bibitem [{\citenamefont {Younis}\ \emph {et~al.}(2020)\citenamefont {Younis},
  \citenamefont {He}, \citenamefont {Lin},\ and\ \citenamefont
  {Huang}}]{Younis2020}%
  \BibitemOpen
  \bibfield  {author} {\bibinfo {author} {\bibfnamefont {M.~R.}\ \bibnamefont
  {Younis}}, \bibinfo {author} {\bibfnamefont {G.}~\bibnamefont {He}}, \bibinfo
  {author} {\bibfnamefont {J.}~\bibnamefont {Lin}}, \ and\ \bibinfo {author}
  {\bibfnamefont {P.}~\bibnamefont {Huang}},\ }\href
  {https://doi.org/10.3389/fchem.2020.00424} {\bibfield  {journal} {\bibinfo
  {journal} {Front. Chem.}\ }\textbf {\bibinfo {volume} {8}} (\bibinfo {year}
  {2020})}\BibitemShut {NoStop}%
\bibitem [{\citenamefont {Gupta}\ \emph {et~al.}(2011)\citenamefont {Gupta},
  \citenamefont {Chaudhary}, \citenamefont {Srivastava}, \citenamefont
  {Sharma}, \citenamefont {Bhardwaj},\ and\ \citenamefont {Chand}}]{Gupta2011}%
  \BibitemOpen
  \bibfield  {author} {\bibinfo {author} {\bibfnamefont {V.}~\bibnamefont
  {Gupta}}, \bibinfo {author} {\bibfnamefont {N.}~\bibnamefont {Chaudhary}},
  \bibinfo {author} {\bibfnamefont {R.}~\bibnamefont {Srivastava}}, \bibinfo
  {author} {\bibfnamefont {G.~D.}\ \bibnamefont {Sharma}}, \bibinfo {author}
  {\bibfnamefont {R.}~\bibnamefont {Bhardwaj}}, \ and\ \bibinfo {author}
  {\bibfnamefont {S.}~\bibnamefont {Chand}},\ }\href {\doibase
  10.1021/ja2036749} {\bibfield  {journal} {\bibinfo  {journal} {J. Am. Chem.
  Soc.}\ }\textbf {\bibinfo {volume} {133}},\ \bibinfo {pages} {9960} (\bibinfo
  {year} {2011})}\BibitemShut {NoStop}%
\bibitem [{\citenamefont {Gao}\ \emph {et~al.}(2014)\citenamefont {Gao},
  \citenamefont {Ding}, \citenamefont {Wang}, \citenamefont {Ruan},
  \citenamefont {Diao}, \citenamefont {Zhang}, \citenamefont {Sun},\ and\
  \citenamefont {Jie}}]{Gao2014}%
  \BibitemOpen
  \bibfield  {author} {\bibinfo {author} {\bibfnamefont {P.}~\bibnamefont
  {Gao}}, \bibinfo {author} {\bibfnamefont {K.}~\bibnamefont {Ding}}, \bibinfo
  {author} {\bibfnamefont {Y.}~\bibnamefont {Wang}}, \bibinfo {author}
  {\bibfnamefont {K.}~\bibnamefont {Ruan}}, \bibinfo {author} {\bibfnamefont
  {S.}~\bibnamefont {Diao}}, \bibinfo {author} {\bibfnamefont {Q.}~\bibnamefont
  {Zhang}}, \bibinfo {author} {\bibfnamefont {B.}~\bibnamefont {Sun}}, \ and\
  \bibinfo {author} {\bibfnamefont {J.}~\bibnamefont {Jie}},\ }\href {\doibase
  10.1021/jp412591k} {\bibfield  {journal} {\bibinfo  {journal} {J. Phys. Chem.
  C}\ }\textbf {\bibinfo {volume} {118}},\ \bibinfo {pages} {5164} (\bibinfo
  {year} {2014})}\BibitemShut {NoStop}%
\bibitem [{\citenamefont {Kim}\ \emph {et~al.}(2020)\citenamefont {Kim},
  \citenamefont {Lee}, \citenamefont {Seo}, \citenamefont {Hong}, \citenamefont
  {Song}, \citenamefont {Lee}, \citenamefont {Ahn}, \citenamefont {Lee},\ and\
  \citenamefont {Song}}]{Kim2020}%
  \BibitemOpen
  \bibfield  {author} {\bibinfo {author} {\bibfnamefont {H.~J.}\ \bibnamefont
  {Kim}}, \bibinfo {author} {\bibfnamefont {C.~K.}\ \bibnamefont {Lee}},
  \bibinfo {author} {\bibfnamefont {J.~G.}\ \bibnamefont {Seo}}, \bibinfo
  {author} {\bibfnamefont {S.~J.}\ \bibnamefont {Hong}}, \bibinfo {author}
  {\bibfnamefont {G.}~\bibnamefont {Song}}, \bibinfo {author} {\bibfnamefont
  {J.}~\bibnamefont {Lee}}, \bibinfo {author} {\bibfnamefont {C.}~\bibnamefont
  {Ahn}}, \bibinfo {author} {\bibfnamefont {D.~J.}\ \bibnamefont {Lee}}, \ and\
  \bibinfo {author} {\bibfnamefont {S.~H.}\ \bibnamefont {Song}},\ }\href
  {\doibase 10.1039/d0ra02257h} {\bibfield  {journal} {\bibinfo  {journal}
  {{RSC} Adv.}\ }\textbf {\bibinfo {volume} {10}},\ \bibinfo {pages} {27418}
  (\bibinfo {year} {2020})}\BibitemShut {NoStop}%
\bibitem [{\citenamefont {Zeng}\ \emph {et~al.}(2018)\citenamefont {Zeng},
  \citenamefont {Chen}, \citenamefont {Tan},\ and\ \citenamefont
  {Xiao}}]{Zeng2018}%
  \BibitemOpen
  \bibfield  {author} {\bibinfo {author} {\bibfnamefont {Z.}~\bibnamefont
  {Zeng}}, \bibinfo {author} {\bibfnamefont {S.}~\bibnamefont {Chen}}, \bibinfo
  {author} {\bibfnamefont {T.~T.~Y.}\ \bibnamefont {Tan}}, \ and\ \bibinfo
  {author} {\bibfnamefont {F.-X.}\ \bibnamefont {Xiao}},\ }\href {\doibase
  10.1016/j.cattod.2018.01.005} {\bibfield  {journal} {\bibinfo  {journal}
  {Catalysis Today}\ }\textbf {\bibinfo {volume} {315}},\ \bibinfo {pages}
  {171} (\bibinfo {year} {2018})}\BibitemShut {NoStop}%
\bibitem [{\citenamefont {Yan}\ \emph {et~al.}(2018)\citenamefont {Yan},
  \citenamefont {Chen}, \citenamefont {Li}, \citenamefont {Tian}, \citenamefont
  {Li}, \citenamefont {Jiang}, \citenamefont {Liu}, \citenamefont {Tian},\ and\
  \citenamefont {Chen}}]{Yan2018}%
  \BibitemOpen
  \bibfield  {author} {\bibinfo {author} {\bibfnamefont {Y.}~\bibnamefont
  {Yan}}, \bibinfo {author} {\bibfnamefont {J.}~\bibnamefont {Chen}}, \bibinfo
  {author} {\bibfnamefont {N.}~\bibnamefont {Li}}, \bibinfo {author}
  {\bibfnamefont {J.}~\bibnamefont {Tian}}, \bibinfo {author} {\bibfnamefont
  {K.}~\bibnamefont {Li}}, \bibinfo {author} {\bibfnamefont {J.}~\bibnamefont
  {Jiang}}, \bibinfo {author} {\bibfnamefont {J.}~\bibnamefont {Liu}}, \bibinfo
  {author} {\bibfnamefont {Q.}~\bibnamefont {Tian}}, \ and\ \bibinfo {author}
  {\bibfnamefont {P.}~\bibnamefont {Chen}},\ }\href {\doibase
  10.1021/acsnano.8b00498} {\bibfield  {journal} {\bibinfo  {journal} {{ACS}
  Nano}\ }\textbf {\bibinfo {volume} {12}},\ \bibinfo {pages} {3523} (\bibinfo
  {year} {2018})}\BibitemShut {NoStop}%
\bibitem [{\citenamefont {Oluwole}\ \emph {et~al.}(2018)\citenamefont
  {Oluwole}, \citenamefont {Nwaji}, \citenamefont {Nene}, \citenamefont
  {Mokone}, \citenamefont {Dube},\ and\ \citenamefont {Nyokong}}]{Oluwole2018}%
  \BibitemOpen
  \bibfield  {author} {\bibinfo {author} {\bibfnamefont {D.~O.}\ \bibnamefont
  {Oluwole}}, \bibinfo {author} {\bibfnamefont {N.}~\bibnamefont {Nwaji}},
  \bibinfo {author} {\bibfnamefont {L.~C.}\ \bibnamefont {Nene}}, \bibinfo
  {author} {\bibfnamefont {L.}~\bibnamefont {Mokone}}, \bibinfo {author}
  {\bibfnamefont {E.}~\bibnamefont {Dube}}, \ and\ \bibinfo {author}
  {\bibfnamefont {T.}~\bibnamefont {Nyokong}},\ }\href {\doibase
  10.1039/c8nj01707g} {\bibfield  {journal} {\bibinfo  {journal} {New J.
  Chem.}\ }\textbf {\bibinfo {volume} {42}},\ \bibinfo {pages} {10124}
  (\bibinfo {year} {2018})}\BibitemShut {NoStop}%
\bibitem [{\citenamefont {Meng}\ \emph {et~al.}(2018)\citenamefont {Meng},
  \citenamefont {Zhao}, \citenamefont {Li},\ and\ \citenamefont
  {Chang}}]{Meng2018}%
  \BibitemOpen
  \bibfield  {author} {\bibinfo {author} {\bibfnamefont {H.-M.}\ \bibnamefont
  {Meng}}, \bibinfo {author} {\bibfnamefont {D.}~\bibnamefont {Zhao}}, \bibinfo
  {author} {\bibfnamefont {N.}~\bibnamefont {Li}}, \ and\ \bibinfo {author}
  {\bibfnamefont {J.}~\bibnamefont {Chang}},\ }\href {\doibase
  10.1039/c8an00677f} {\bibfield  {journal} {\bibinfo  {journal} {Analyst}\
  }\textbf {\bibinfo {volume} {143}},\ \bibinfo {pages} {4967} (\bibinfo {year}
  {2018})}\BibitemShut {NoStop}%
\bibitem [{\citenamefont {Kuo}\ \emph {et~al.}(2020)\citenamefont {Kuo},
  \citenamefont {Shen}, \citenamefont {Chang}, \citenamefont {Kao},
  \citenamefont {Lin}, \citenamefont {Wang},\ and\ \citenamefont
  {Wu}}]{Kuo2020}%
  \BibitemOpen
  \bibfield  {author} {\bibinfo {author} {\bibfnamefont {W.-S.}\ \bibnamefont
  {Kuo}}, \bibinfo {author} {\bibfnamefont {X.-C.}\ \bibnamefont {Shen}},
  \bibinfo {author} {\bibfnamefont {C.-Y.}\ \bibnamefont {Chang}}, \bibinfo
  {author} {\bibfnamefont {H.-F.}\ \bibnamefont {Kao}}, \bibinfo {author}
  {\bibfnamefont {S.-H.}\ \bibnamefont {Lin}}, \bibinfo {author} {\bibfnamefont
  {J.-Y.}\ \bibnamefont {Wang}}, \ and\ \bibinfo {author} {\bibfnamefont
  {P.-C.}\ \bibnamefont {Wu}},\ }\href {\doibase 10.1021/acsnano.0c03915}
  {\bibfield  {journal} {\bibinfo  {journal} {{ACS} Nano}\ }\textbf {\bibinfo
  {volume} {14}},\ \bibinfo {pages} {11502} (\bibinfo {year}
  {2020})}\BibitemShut {NoStop}%
\bibitem [{\citenamefont {Trauzettel}\ \emph {et~al.}(2007)\citenamefont
  {Trauzettel}, \citenamefont {Bulaev}, \citenamefont {Loss},\ and\
  \citenamefont {Burkard}}]{Trauzettel2007}%
  \BibitemOpen
  \bibfield  {author} {\bibinfo {author} {\bibfnamefont {B.}~\bibnamefont
  {Trauzettel}}, \bibinfo {author} {\bibfnamefont {D.~V.}\ \bibnamefont
  {Bulaev}}, \bibinfo {author} {\bibfnamefont {D.}~\bibnamefont {Loss}}, \ and\
  \bibinfo {author} {\bibfnamefont {G.}~\bibnamefont {Burkard}},\ }\href
  {\doibase 10.1038/nphys544} {\bibfield  {journal} {\bibinfo  {journal} {Nat.
  Phys.}\ }\textbf {\bibinfo {volume} {3}},\ \bibinfo {pages} {192} (\bibinfo
  {year} {2007})}\BibitemShut {NoStop}%
\bibitem [{\citenamefont {Eich}\ \emph {et~al.}(2018)\citenamefont {Eich} \emph
  {et~al.}}]{Eich_2018}%
  \BibitemOpen
  \bibfield  {author} {\bibinfo {author} {\bibfnamefont {M.}~\bibnamefont
  {Eich}} \emph {et~al.},\ }\href {\doibase 10.1103/PhysRevX.8.031023}
  {\bibfield  {journal} {\bibinfo  {journal} {Phys. Rev. X}\ }\textbf {\bibinfo
  {volume} {8}},\ \bibinfo {pages} {031023} (\bibinfo {year}
  {2018})}\BibitemShut {NoStop}%
\bibitem [{\citenamefont {Son}\ \emph {et~al.}(2006)\citenamefont {Son},
  \citenamefont {Cohen},\ and\ \citenamefont {Louie}}]{Louie2006}%
  \BibitemOpen
  \bibfield  {author} {\bibinfo {author} {\bibfnamefont {Y.-W.}\ \bibnamefont
  {Son}}, \bibinfo {author} {\bibfnamefont {M.~L.}\ \bibnamefont {Cohen}}, \
  and\ \bibinfo {author} {\bibfnamefont {S.~G.}\ \bibnamefont {Louie}},\ }\href
  {\doibase 10.1103/PhysRevLett.97.216803} {\bibfield  {journal} {\bibinfo
  {journal} {Phys. Rev. Lett.}\ }\textbf {\bibinfo {volume} {97}},\ \bibinfo
  {pages} {216803} (\bibinfo {year} {2006})}\BibitemShut {NoStop}%
\bibitem [{\citenamefont {Zhang}\ \emph {et~al.}(1998)\citenamefont {Zhang},
  \citenamefont {Bertran},\ and\ \citenamefont {Lee}}]{Zhang_1998}%
  \BibitemOpen
  \bibfield  {author} {\bibinfo {author} {\bibfnamefont {R.}~\bibnamefont
  {Zhang}}, \bibinfo {author} {\bibfnamefont {E.}~\bibnamefont {Bertran}}, \
  and\ \bibinfo {author} {\bibfnamefont {S.}~\bibnamefont {Lee}},\ }\href@noop
  {} {\bibfield  {journal} {\bibinfo  {journal} {Diam. Relat. Mater.}\ }\textbf
  {\bibinfo {volume} {7}},\ \bibinfo {pages} {1663} (\bibinfo {year}
  {1998})}\BibitemShut {NoStop}%
\bibitem [{\citenamefont {Ritter}\ and\ \citenamefont
  {Lyding}(2009)}]{Ritter_2009}%
  \BibitemOpen
  \bibfield  {author} {\bibinfo {author} {\bibfnamefont {K.~A.}\ \bibnamefont
  {Ritter}}\ and\ \bibinfo {author} {\bibfnamefont {J.~W.}\ \bibnamefont
  {Lyding}},\ }\href {\doibase 10.1038/nmat2378} {\bibfield  {journal}
  {\bibinfo  {journal} {Nat. Mater.}\ }\textbf {\bibinfo {volume} {8}},\
  \bibinfo {pages} {235} (\bibinfo {year} {2009})}\BibitemShut {NoStop}%
\bibitem [{\citenamefont {Magda}\ \emph {et~al.}(2014)\citenamefont {Magda},
  \citenamefont {Jin}, \citenamefont {Hagym{\'{a}}si}, \citenamefont
  {Vancs{\'{o}}}, \citenamefont {Osv{\'{a}}th}, \citenamefont {Nemes-Incze},
  \citenamefont {Hwang}, \citenamefont {Bir{\'{o}}},\ and\ \citenamefont
  {Tapaszt{\'{o}}}}]{Magda2014}%
  \BibitemOpen
  \bibfield  {author} {\bibinfo {author} {\bibfnamefont {G.~Z.}\ \bibnamefont
  {Magda}}, \bibinfo {author} {\bibfnamefont {X.}~\bibnamefont {Jin}}, \bibinfo
  {author} {\bibfnamefont {I.}~\bibnamefont {Hagym{\'{a}}si}}, \bibinfo
  {author} {\bibfnamefont {P.}~\bibnamefont {Vancs{\'{o}}}}, \bibinfo {author}
  {\bibfnamefont {Z.}~\bibnamefont {Osv{\'{a}}th}}, \bibinfo {author}
  {\bibfnamefont {P.}~\bibnamefont {Nemes-Incze}}, \bibinfo {author}
  {\bibfnamefont {C.}~\bibnamefont {Hwang}}, \bibinfo {author} {\bibfnamefont
  {L.~P.}\ \bibnamefont {Bir{\'{o}}}}, \ and\ \bibinfo {author} {\bibfnamefont
  {L.}~\bibnamefont {Tapaszt{\'{o}}}},\ }\href {\doibase 10.1038/nature13831}
  {\bibfield  {journal} {\bibinfo  {journal} {Nature}\ }\textbf {\bibinfo
  {volume} {514}},\ \bibinfo {pages} {608} (\bibinfo {year}
  {2014})}\BibitemShut {NoStop}%
\bibitem [{\citenamefont {Cort{\'{e}}s-del R{\'{\i}}o}\ \emph
  {et~al.}(2020)\citenamefont {Cort{\'{e}}s-del R{\'{\i}}o}, \citenamefont
  {Mallet}, \citenamefont {Gonz{\'{a}}lez-Herrero}, \citenamefont {Lado},
  \citenamefont {Fern{\'{a}}ndez-Rossier}, \citenamefont
  {G{\'{o}}mez-Rodr{\'{\i}}guez}, \citenamefont {Veuillen},\ and\ \citenamefont
  {Brihuega}}]{CortsdelRo2020}%
  \BibitemOpen
  \bibfield  {author} {\bibinfo {author} {\bibfnamefont {E.}~\bibnamefont
  {Cort{\'{e}}s-del R{\'{\i}}o}}, \bibinfo {author} {\bibfnamefont
  {P.}~\bibnamefont {Mallet}}, \bibinfo {author} {\bibfnamefont
  {H.}~\bibnamefont {Gonz{\'{a}}lez-Herrero}}, \bibinfo {author} {\bibfnamefont
  {J.~L.}\ \bibnamefont {Lado}}, \bibinfo {author} {\bibfnamefont
  {J.}~\bibnamefont {Fern{\'{a}}ndez-Rossier}}, \bibinfo {author}
  {\bibfnamefont {J.~M.}\ \bibnamefont {G{\'{o}}mez-Rodr{\'{\i}}guez}},
  \bibinfo {author} {\bibfnamefont {J.-Y.}\ \bibnamefont {Veuillen}}, \ and\
  \bibinfo {author} {\bibfnamefont {I.}~\bibnamefont {Brihuega}},\ }\href
  {\doibase 10.1002/adma.202001119} {\bibfield  {journal} {\bibinfo  {journal}
  {Adv. Mater.}\ }\textbf {\bibinfo {volume} {32}},\ \bibinfo {pages} {2001119}
  (\bibinfo {year} {2020})}\BibitemShut {NoStop}%
\bibitem [{\citenamefont {Zhang}\ \emph {et~al.}(2008)\citenamefont {Zhang},
  \citenamefont {Chang},\ and\ \citenamefont {Peeters}}]{Zhang_2008}%
  \BibitemOpen
  \bibfield  {author} {\bibinfo {author} {\bibfnamefont {Z.~Z.}\ \bibnamefont
  {Zhang}}, \bibinfo {author} {\bibfnamefont {K.}~\bibnamefont {Chang}}, \ and\
  \bibinfo {author} {\bibfnamefont {F.~M.}\ \bibnamefont {Peeters}},\ }\href
  {\doibase 10.1103/PhysRevB.77.235411} {\bibfield  {journal} {\bibinfo
  {journal} {Phys. Rev. B}\ }\textbf {\bibinfo {volume} {77}},\ \bibinfo
  {pages} {235411} (\bibinfo {year} {2008})}\BibitemShut {NoStop}%
\bibitem [{\citenamefont {Saleem}\ \emph {et~al.}(2019)\citenamefont {Saleem},
  \citenamefont {Baldo}, \citenamefont {Delgado}, \citenamefont {Szulakowska},\
  and\ \citenamefont {Hawrylak}}]{Saleem_2019}%
  \BibitemOpen
  \bibfield  {author} {\bibinfo {author} {\bibfnamefont {Y.}~\bibnamefont
  {Saleem}}, \bibinfo {author} {\bibfnamefont {L.~N.}\ \bibnamefont {Baldo}},
  \bibinfo {author} {\bibfnamefont {A.}~\bibnamefont {Delgado}}, \bibinfo
  {author} {\bibfnamefont {L.}~\bibnamefont {Szulakowska}}, \ and\ \bibinfo
  {author} {\bibfnamefont {P.}~\bibnamefont {Hawrylak}},\ }\href {\doibase
  10.1088/1361-648x/ab0b31} {\bibfield  {journal} {\bibinfo  {journal} {J.
  Phys.: Condens. Matter}\ }\textbf {\bibinfo {volume} {31}},\ \bibinfo {pages}
  {305503} (\bibinfo {year} {2019})}\BibitemShut {NoStop}%
\bibitem [{\citenamefont {Nakada}\ \emph {et~al.}(1996)\citenamefont {Nakada},
  \citenamefont {Fujita}, \citenamefont {Dresselhaus},\ and\ \citenamefont
  {Dresselhaus}}]{Kyoko1996}%
  \BibitemOpen
  \bibfield  {author} {\bibinfo {author} {\bibfnamefont {K.}~\bibnamefont
  {Nakada}}, \bibinfo {author} {\bibfnamefont {M.}~\bibnamefont {Fujita}},
  \bibinfo {author} {\bibfnamefont {G.}~\bibnamefont {Dresselhaus}}, \ and\
  \bibinfo {author} {\bibfnamefont {M.~S.}\ \bibnamefont {Dresselhaus}},\
  }\href {\doibase 10.1103/PhysRevB.54.17954} {\bibfield  {journal} {\bibinfo
  {journal} {Phys. Rev. B}\ }\textbf {\bibinfo {volume} {54}},\ \bibinfo
  {pages} {17954} (\bibinfo {year} {1996})}\BibitemShut {NoStop}%
\bibitem [{\citenamefont {Yan}\ \emph {et~al.}(2010)\citenamefont {Yan},
  \citenamefont {Cui},\ and\ \citenamefont {Li}}]{Yan2010}%
  \BibitemOpen
  \bibfield  {author} {\bibinfo {author} {\bibfnamefont {X.}~\bibnamefont
  {Yan}}, \bibinfo {author} {\bibfnamefont {X.}~\bibnamefont {Cui}}, \ and\
  \bibinfo {author} {\bibfnamefont {L.-S.}\ \bibnamefont {Li}},\ }\href
  {\doibase 10.1021/ja1009376} {\bibfield  {journal} {\bibinfo  {journal} {J.
  Am. Chem. Soc.}\ }\textbf {\bibinfo {volume} {132}},\ \bibinfo {pages} {5944}
  (\bibinfo {year} {2010})}\BibitemShut {NoStop}%
\bibitem [{\citenamefont {Pan}\ \emph {et~al.}(2010)\citenamefont {Pan},
  \citenamefont {Zhang}, \citenamefont {Li},\ and\ \citenamefont
  {Wu}}]{Pan2010}%
  \BibitemOpen
  \bibfield  {author} {\bibinfo {author} {\bibfnamefont {D.}~\bibnamefont
  {Pan}}, \bibinfo {author} {\bibfnamefont {J.}~\bibnamefont {Zhang}}, \bibinfo
  {author} {\bibfnamefont {Z.}~\bibnamefont {Li}}, \ and\ \bibinfo {author}
  {\bibfnamefont {M.}~\bibnamefont {Wu}},\ }\href {\doibase
  10.1002/adma.200902825} {\bibfield  {journal} {\bibinfo  {journal} {Adv.
  Mater.}\ }\textbf {\bibinfo {volume} {22}},\ \bibinfo {pages} {734} (\bibinfo
  {year} {2010})}\BibitemShut {NoStop}%
\bibitem [{\citenamefont {Wimmer}\ \emph {et~al.}(2010)\citenamefont {Wimmer},
  \citenamefont {Akhmerov},\ and\ \citenamefont {Guinea}}]{Wimmer2010}%
  \BibitemOpen
  \bibfield  {author} {\bibinfo {author} {\bibfnamefont {M.}~\bibnamefont
  {Wimmer}}, \bibinfo {author} {\bibfnamefont {A.~R.}\ \bibnamefont
  {Akhmerov}}, \ and\ \bibinfo {author} {\bibfnamefont {F.}~\bibnamefont
  {Guinea}},\ }\href {\doibase 10.1103/PhysRevB.82.045409} {\bibfield
  {journal} {\bibinfo  {journal} {Phys. Rev. B}\ }\textbf {\bibinfo {volume}
  {82}},\ \bibinfo {pages} {045409} (\bibinfo {year} {2010})}\BibitemShut
  {NoStop}%
\bibitem [{\citenamefont {Ozfidan}\ \emph {et~al.}(2015)\citenamefont
  {Ozfidan}, \citenamefont {Korkusinski},\ and\ \citenamefont
  {Hawrylak}}]{Ozfidan2015}%
  \BibitemOpen
  \bibfield  {author} {\bibinfo {author} {\bibfnamefont {I.}~\bibnamefont
  {Ozfidan}}, \bibinfo {author} {\bibfnamefont {M.}~\bibnamefont
  {Korkusinski}}, \ and\ \bibinfo {author} {\bibfnamefont {P.}~\bibnamefont
  {Hawrylak}},\ }\href {\doibase 10.1002/pssr.201510251} {\bibfield  {journal}
  {\bibinfo  {journal} {Phys. Status Solidi RRL}\ }\textbf {\bibinfo {volume}
  {10}},\ \bibinfo {pages} {13} (\bibinfo {year} {2015})}\BibitemShut {NoStop}%
\bibitem [{\citenamefont {Li}\ \emph {et~al.}(2011)\citenamefont {Li},
  \citenamefont {Zhao}, \citenamefont {Cheng}, \citenamefont {Hu},
  \citenamefont {Shi}, \citenamefont {Dai},\ and\ \citenamefont {Qu}}]{Li2011}%
  \BibitemOpen
  \bibfield  {author} {\bibinfo {author} {\bibfnamefont {Y.}~\bibnamefont
  {Li}}, \bibinfo {author} {\bibfnamefont {Y.}~\bibnamefont {Zhao}}, \bibinfo
  {author} {\bibfnamefont {H.}~\bibnamefont {Cheng}}, \bibinfo {author}
  {\bibfnamefont {Y.}~\bibnamefont {Hu}}, \bibinfo {author} {\bibfnamefont
  {G.}~\bibnamefont {Shi}}, \bibinfo {author} {\bibfnamefont {L.}~\bibnamefont
  {Dai}}, \ and\ \bibinfo {author} {\bibfnamefont {L.}~\bibnamefont {Qu}},\
  }\href {\doibase 10.1021/ja206030c} {\bibfield  {journal} {\bibinfo
  {journal} {J. Am. Chem. Soc.}\ }\textbf {\bibinfo {volume} {134}},\ \bibinfo
  {pages} {15} (\bibinfo {year} {2011})}\BibitemShut {NoStop}%
\bibitem [{\citenamefont {Yan}\ and\ \citenamefont {shi Li}(2011)}]{Yan2011}%
  \BibitemOpen
  \bibfield  {author} {\bibinfo {author} {\bibfnamefont {X.}~\bibnamefont
  {Yan}}\ and\ \bibinfo {author} {\bibfnamefont {L.}~\bibnamefont {shi Li}},\
  }\href {\doibase 10.1039/c0jm02827d} {\bibfield  {journal} {\bibinfo
  {journal} {J. Mater. Chem.}\ }\textbf {\bibinfo {volume} {21}},\ \bibinfo
  {pages} {3295} (\bibinfo {year} {2011})}\BibitemShut {NoStop}%
\bibitem [{\citenamefont {Qian}\ \emph {et~al.}(2013)\citenamefont {Qian},
  \citenamefont {Ma}, \citenamefont {Shan}, \citenamefont {Shao}, \citenamefont
  {Zhou}, \citenamefont {Chen},\ and\ \citenamefont {Feng}}]{Qian2013}%
  \BibitemOpen
  \bibfield  {author} {\bibinfo {author} {\bibfnamefont {Z.}~\bibnamefont
  {Qian}}, \bibinfo {author} {\bibfnamefont {J.}~\bibnamefont {Ma}}, \bibinfo
  {author} {\bibfnamefont {X.}~\bibnamefont {Shan}}, \bibinfo {author}
  {\bibfnamefont {L.}~\bibnamefont {Shao}}, \bibinfo {author} {\bibfnamefont
  {J.}~\bibnamefont {Zhou}}, \bibinfo {author} {\bibfnamefont {J.}~\bibnamefont
  {Chen}}, \ and\ \bibinfo {author} {\bibfnamefont {H.}~\bibnamefont {Feng}},\
  }\href {\doibase 10.1039/c3ra42066c} {\bibfield  {journal} {\bibinfo
  {journal} {{RSC} Adv.}\ }\textbf {\bibinfo {volume} {3}},\ \bibinfo {pages}
  {14571} (\bibinfo {year} {2013})}\BibitemShut {NoStop}%
\bibitem [{\citenamefont {Wang}\ \emph
  {et~al.}(2016{\natexlab{a}})\citenamefont {Wang}, \citenamefont {Cole},
  \citenamefont {Zhao},\ and\ \citenamefont {Li}}]{Wang2016}%
  \BibitemOpen
  \bibfield  {author} {\bibinfo {author} {\bibfnamefont {S.}~\bibnamefont
  {Wang}}, \bibinfo {author} {\bibfnamefont {I.~S.}\ \bibnamefont {Cole}},
  \bibinfo {author} {\bibfnamefont {D.}~\bibnamefont {Zhao}}, \ and\ \bibinfo
  {author} {\bibfnamefont {Q.}~\bibnamefont {Li}},\ }\href {\doibase
  10.1039/c5nr07042b} {\bibfield  {journal} {\bibinfo  {journal} {Nanoscale}\
  }\textbf {\bibinfo {volume} {8}},\ \bibinfo {pages} {7449} (\bibinfo {year}
  {2016}{\natexlab{a}})}\BibitemShut {NoStop}%
\bibitem [{\citenamefont {Wang}\ \emph
  {et~al.}(2016{\natexlab{b}})\citenamefont {Wang}, \citenamefont {Dong},
  \citenamefont {Sun}, \citenamefont {Chen}, \citenamefont {Wang},
  \citenamefont {Wang},\ and\ \citenamefont {Dong}}]{Wang2016_2}%
  \BibitemOpen
  \bibfield  {author} {\bibinfo {author} {\bibfnamefont {K.}~\bibnamefont
  {Wang}}, \bibinfo {author} {\bibfnamefont {J.}~\bibnamefont {Dong}}, \bibinfo
  {author} {\bibfnamefont {L.}~\bibnamefont {Sun}}, \bibinfo {author}
  {\bibfnamefont {H.}~\bibnamefont {Chen}}, \bibinfo {author} {\bibfnamefont
  {Y.}~\bibnamefont {Wang}}, \bibinfo {author} {\bibfnamefont {C.}~\bibnamefont
  {Wang}}, \ and\ \bibinfo {author} {\bibfnamefont {L.}~\bibnamefont {Dong}},\
  }\href {\doibase 10.1039/c6ra19673j} {\bibfield  {journal} {\bibinfo
  {journal} {{RSC} Adv.}\ }\textbf {\bibinfo {volume} {6}},\ \bibinfo {pages}
  {91225} (\bibinfo {year} {2016}{\natexlab{b}})}\BibitemShut {NoStop}%
\bibitem [{\citenamefont {Qian}\ \emph {et~al.}(2016)\citenamefont {Qian},
  \citenamefont {Li}, \citenamefont {Tang}, \citenamefont {Lai}, \citenamefont
  {Lu},\ and\ \citenamefont {Lau}}]{Qian2016}%
  \BibitemOpen
  \bibfield  {author} {\bibinfo {author} {\bibfnamefont {F.}~\bibnamefont
  {Qian}}, \bibinfo {author} {\bibfnamefont {X.}~\bibnamefont {Li}}, \bibinfo
  {author} {\bibfnamefont {L.}~\bibnamefont {Tang}}, \bibinfo {author}
  {\bibfnamefont {S.~K.}\ \bibnamefont {Lai}}, \bibinfo {author} {\bibfnamefont
  {C.}~\bibnamefont {Lu}}, \ and\ \bibinfo {author} {\bibfnamefont {S.~P.}\
  \bibnamefont {Lau}},\ }\href {\doibase 10.1063/1.4959906} {\bibfield
  {journal} {\bibinfo  {journal} {{AIP} Adv.}\ }\textbf {\bibinfo {volume}
  {6}},\ \bibinfo {pages} {075116} (\bibinfo {year} {2016})}\BibitemShut
  {NoStop}%
\bibitem [{\citenamefont {Bayoumy}\ \emph {et~al.}(2019)\citenamefont
  {Bayoumy}, \citenamefont {Refaat}, \citenamefont {Yahia}, \citenamefont
  {Zahran}, \citenamefont {Elhaes}, \citenamefont {Ibrahim},\ and\
  \citenamefont {Shkir}}]{Bayoumy2019}%
  \BibitemOpen
  \bibfield  {author} {\bibinfo {author} {\bibfnamefont {A.~M.}\ \bibnamefont
  {Bayoumy}}, \bibinfo {author} {\bibfnamefont {A.}~\bibnamefont {Refaat}},
  \bibinfo {author} {\bibfnamefont {I.~S.}\ \bibnamefont {Yahia}}, \bibinfo
  {author} {\bibfnamefont {H.~Y.}\ \bibnamefont {Zahran}}, \bibinfo {author}
  {\bibfnamefont {H.}~\bibnamefont {Elhaes}}, \bibinfo {author} {\bibfnamefont
  {M.~A.}\ \bibnamefont {Ibrahim}}, \ and\ \bibinfo {author} {\bibfnamefont
  {M.}~\bibnamefont {Shkir}},\ }\href
  {https://doi.org/10.1007/s11082-019-2134-z} {\bibfield  {journal} {\bibinfo
  {journal} {Opt. Quantum Electron.}\ }\textbf {\bibinfo {volume} {52}}
  (\bibinfo {year} {2019})}\BibitemShut {NoStop}%
\bibitem [{\citenamefont {Kadian}\ \emph {et~al.}(2019)\citenamefont {Kadian},
  \citenamefont {Manik}, \citenamefont {Kalkal}, \citenamefont {Singh},\ and\
  \citenamefont {Chauhan}}]{Kadian_2019}%
  \BibitemOpen
  \bibfield  {author} {\bibinfo {author} {\bibfnamefont {S.}~\bibnamefont
  {Kadian}}, \bibinfo {author} {\bibfnamefont {G.}~\bibnamefont {Manik}},
  \bibinfo {author} {\bibfnamefont {A.}~\bibnamefont {Kalkal}}, \bibinfo
  {author} {\bibfnamefont {M.}~\bibnamefont {Singh}}, \ and\ \bibinfo {author}
  {\bibfnamefont {R.~P.}\ \bibnamefont {Chauhan}},\ }\href {\doibase
  10.1088/1361-6528/ab3566} {\bibfield  {journal} {\bibinfo  {journal}
  {Nanotechnology}\ }\textbf {\bibinfo {volume} {30}},\ \bibinfo {pages}
  {435704} (\bibinfo {year} {2019})}\BibitemShut {NoStop}%
\bibitem [{\citenamefont {Ci}\ \emph {et~al.}(2010)\citenamefont {Ci} \emph
  {et~al.}}]{Ci2010}%
  \BibitemOpen
  \bibfield  {author} {\bibinfo {author} {\bibfnamefont {L.}~\bibnamefont {Ci}}
  \emph {et~al.},\ }\href {\doibase 10.1038/nmat2711} {\bibfield  {journal}
  {\bibinfo  {journal} {Nat. Mater.}\ }\textbf {\bibinfo {volume} {9}},\
  \bibinfo {pages} {430} (\bibinfo {year} {2010})}\BibitemShut {NoStop}%
\bibitem [{\citenamefont {Peng}\ \emph {et~al.}(2013)\citenamefont {Peng},
  \citenamefont {Wang}, \citenamefont {Zhang}, \citenamefont {Jiang},
  \citenamefont {Shi},\ and\ \citenamefont {Zhu}}]{Peng2013}%
  \BibitemOpen
  \bibfield  {author} {\bibinfo {author} {\bibfnamefont {J.}~\bibnamefont
  {Peng}}, \bibinfo {author} {\bibfnamefont {S.}~\bibnamefont {Wang}}, \bibinfo
  {author} {\bibfnamefont {P.-H.}\ \bibnamefont {Zhang}}, \bibinfo {author}
  {\bibfnamefont {L.-P.}\ \bibnamefont {Jiang}}, \bibinfo {author}
  {\bibfnamefont {J.-J.}\ \bibnamefont {Shi}}, \ and\ \bibinfo {author}
  {\bibfnamefont {J.-J.}\ \bibnamefont {Zhu}},\ }\href {\doibase
  10.1166/jbn.2013.1663} {\bibfield  {journal} {\bibinfo  {journal} {J. Biomed.
  Nanotechnol.}\ }\textbf {\bibinfo {volume} {9}},\ \bibinfo {pages} {1679}
  (\bibinfo {year} {2013})}\BibitemShut {NoStop}%
\bibitem [{\citenamefont {Kang}\ \emph {et~al.}(2013)\citenamefont {Kang},
  \citenamefont {Chu}, \citenamefont {Zhang}, \citenamefont {Li}, \citenamefont
  {Jiang}, \citenamefont {Cheng},\ and\ \citenamefont {Li}}]{Kang2013}%
  \BibitemOpen
  \bibfield  {author} {\bibinfo {author} {\bibfnamefont {Y.}~\bibnamefont
  {Kang}}, \bibinfo {author} {\bibfnamefont {Z.}~\bibnamefont {Chu}}, \bibinfo
  {author} {\bibfnamefont {D.}~\bibnamefont {Zhang}}, \bibinfo {author}
  {\bibfnamefont {G.}~\bibnamefont {Li}}, \bibinfo {author} {\bibfnamefont
  {Z.}~\bibnamefont {Jiang}}, \bibinfo {author} {\bibfnamefont
  {H.}~\bibnamefont {Cheng}}, \ and\ \bibinfo {author} {\bibfnamefont
  {X.}~\bibnamefont {Li}},\ }\href {\doibase 10.1016/j.carbon.2013.04.085}
  {\bibfield  {journal} {\bibinfo  {journal} {Carbon}\ }\textbf {\bibinfo
  {volume} {61}},\ \bibinfo {pages} {200} (\bibinfo {year} {2013})}\BibitemShut
  {NoStop}%
\bibitem [{\citenamefont {Liu}\ \emph {et~al.}(2013)\citenamefont {Liu} \emph
  {et~al.}}]{Liu2013}%
  \BibitemOpen
  \bibfield  {author} {\bibinfo {author} {\bibfnamefont {Z.}~\bibnamefont
  {Liu}} \emph {et~al.},\ }\href {\doibase 10.1038/nnano.2012.256} {\bibfield
  {journal} {\bibinfo  {journal} {Nat. Nanotechnol.}\ }\textbf {\bibinfo
  {volume} {8}},\ \bibinfo {pages} {119} (\bibinfo {year} {2013})}\BibitemShut
  {NoStop}%
\bibitem [{\citenamefont {Liu}\ \emph {et~al.}(2014)\citenamefont {Liu},
  \citenamefont {Park}, \citenamefont {Siegel}, \citenamefont {McCarty},
  \citenamefont {Clark}, \citenamefont {Deng}, \citenamefont {Basile},
  \citenamefont {Idrobo}, \citenamefont {Li},\ and\ \citenamefont
  {Gu}}]{Liu2014}%
  \BibitemOpen
  \bibfield  {author} {\bibinfo {author} {\bibfnamefont {L.}~\bibnamefont
  {Liu}}, \bibinfo {author} {\bibfnamefont {J.}~\bibnamefont {Park}}, \bibinfo
  {author} {\bibfnamefont {D.~A.}\ \bibnamefont {Siegel}}, \bibinfo {author}
  {\bibfnamefont {K.~F.}\ \bibnamefont {McCarty}}, \bibinfo {author}
  {\bibfnamefont {K.~W.}\ \bibnamefont {Clark}}, \bibinfo {author}
  {\bibfnamefont {W.}~\bibnamefont {Deng}}, \bibinfo {author} {\bibfnamefont
  {L.}~\bibnamefont {Basile}}, \bibinfo {author} {\bibfnamefont {J.~C.}\
  \bibnamefont {Idrobo}}, \bibinfo {author} {\bibfnamefont {A.-P.}\
  \bibnamefont {Li}}, \ and\ \bibinfo {author} {\bibfnamefont {G.}~\bibnamefont
  {Gu}},\ }\href {\doibase 10.1126/science.1246137} {\bibfield  {journal}
  {\bibinfo  {journal} {Science}\ }\textbf {\bibinfo {volume} {343}},\ \bibinfo
  {pages} {163} (\bibinfo {year} {2014})}\BibitemShut {NoStop}%
\bibitem [{\citenamefont {Kim}\ \emph {et~al.}(2015)\citenamefont {Kim},
  \citenamefont {Lim}, \citenamefont {Ma}, \citenamefont {Jang}, \citenamefont
  {Ryu}, \citenamefont {Jung}, \citenamefont {Shin}, \citenamefont {Lee},\ and\
  \citenamefont {Shin}}]{Kim2015}%
  \BibitemOpen
  \bibfield  {author} {\bibinfo {author} {\bibfnamefont {G.}~\bibnamefont
  {Kim}}, \bibinfo {author} {\bibfnamefont {H.}~\bibnamefont {Lim}}, \bibinfo
  {author} {\bibfnamefont {K.~Y.}\ \bibnamefont {Ma}}, \bibinfo {author}
  {\bibfnamefont {A.-R.}\ \bibnamefont {Jang}}, \bibinfo {author}
  {\bibfnamefont {G.~H.}\ \bibnamefont {Ryu}}, \bibinfo {author} {\bibfnamefont
  {M.}~\bibnamefont {Jung}}, \bibinfo {author} {\bibfnamefont {H.-J.}\
  \bibnamefont {Shin}}, \bibinfo {author} {\bibfnamefont {Z.}~\bibnamefont
  {Lee}}, \ and\ \bibinfo {author} {\bibfnamefont {H.~S.}\ \bibnamefont
  {Shin}},\ }\href {\doibase 10.1021/acs.nanolett.5b01704} {\bibfield
  {journal} {\bibinfo  {journal} {Nano Lett.}\ }\textbf {\bibinfo {volume}
  {15}},\ \bibinfo {pages} {4769} (\bibinfo {year} {2015})}\BibitemShut
  {NoStop}%
\bibitem [{\citenamefont {Ghahari}\ \emph {et~al.}(2017)\citenamefont {Ghahari}
  \emph {et~al.}}]{Ghahari2017}%
  \BibitemOpen
  \bibfield  {author} {\bibinfo {author} {\bibfnamefont {F.}~\bibnamefont
  {Ghahari}} \emph {et~al.},\ }\href {\doibase 10.1126/science.aal0212}
  {\bibfield  {journal} {\bibinfo  {journal} {Science}\ }\textbf {\bibinfo
  {volume} {356}},\ \bibinfo {pages} {845} (\bibinfo {year}
  {2017})}\BibitemShut {NoStop}%
\bibitem [{\citenamefont {Chen}\ \emph {et~al.}(2019)\citenamefont {Chen} \emph
  {et~al.}}]{Chen_2019}%
  \BibitemOpen
  \bibfield  {author} {\bibinfo {author} {\bibfnamefont {D.}~\bibnamefont
  {Chen}} \emph {et~al.},\ }\href {\doibase 10.1039/C9NR00412B} {\bibfield
  {journal} {\bibinfo  {journal} {Nanoscale}\ }\textbf {\bibinfo {volume}
  {11}},\ \bibinfo {pages} {4226} (\bibinfo {year} {2019})}\BibitemShut
  {NoStop}%
\bibitem [{\citenamefont {Koskinen}\ \emph {et~al.}(2008)\citenamefont
  {Koskinen}, \citenamefont {Malola},\ and\ \citenamefont
  {H\"akkinen}}]{Pekka2008}%
  \BibitemOpen
  \bibfield  {author} {\bibinfo {author} {\bibfnamefont {P.}~\bibnamefont
  {Koskinen}}, \bibinfo {author} {\bibfnamefont {S.}~\bibnamefont {Malola}}, \
  and\ \bibinfo {author} {\bibfnamefont {H.}~\bibnamefont {H\"akkinen}},\
  }\href {\doibase 10.1103/PhysRevLett.101.115502} {\bibfield  {journal}
  {\bibinfo  {journal} {Phys. Rev. Lett.}\ }\textbf {\bibinfo {volume} {101}},\
  \bibinfo {pages} {115502} (\bibinfo {year} {2008})}\BibitemShut {NoStop}%
\bibitem [{\citenamefont {Girit}\ \emph {et~al.}(2009)\citenamefont {Girit},
  \citenamefont {Meyer}, \citenamefont {Erni}, \citenamefont {Rossell},
  \citenamefont {Kisielowski}, \citenamefont {Yang}, \citenamefont {Park},
  \citenamefont {Crommie}, \citenamefont {Cohen}, \citenamefont {Louie},\ and\
  \citenamefont {Zettl}}]{Girit2009}%
  \BibitemOpen
  \bibfield  {author} {\bibinfo {author} {\bibfnamefont {C.~O.}\ \bibnamefont
  {Girit}}, \bibinfo {author} {\bibfnamefont {J.~C.}\ \bibnamefont {Meyer}},
  \bibinfo {author} {\bibfnamefont {R.}~\bibnamefont {Erni}}, \bibinfo {author}
  {\bibfnamefont {M.~D.}\ \bibnamefont {Rossell}}, \bibinfo {author}
  {\bibfnamefont {C.}~\bibnamefont {Kisielowski}}, \bibinfo {author}
  {\bibfnamefont {L.}~\bibnamefont {Yang}}, \bibinfo {author} {\bibfnamefont
  {C.-H.}\ \bibnamefont {Park}}, \bibinfo {author} {\bibfnamefont {M.~F.}\
  \bibnamefont {Crommie}}, \bibinfo {author} {\bibfnamefont {M.~L.}\
  \bibnamefont {Cohen}}, \bibinfo {author} {\bibfnamefont {S.~G.}\ \bibnamefont
  {Louie}}, \ and\ \bibinfo {author} {\bibfnamefont {A.}~\bibnamefont
  {Zettl}},\ }\href {\doibase 10.1126/science.1166999} {\bibfield  {journal}
  {\bibinfo  {journal} {Science}\ }\textbf {\bibinfo {volume} {323}},\ \bibinfo
  {pages} {1705} (\bibinfo {year} {2009})}\BibitemShut {NoStop}%
\bibitem [{\citenamefont {Kim}\ \emph {et~al.}(2013)\citenamefont {Kim},
  \citenamefont {Coh}, \citenamefont {Kisielowski}, \citenamefont {Crommie},
  \citenamefont {Louie}, \citenamefont {Cohen},\ and\ \citenamefont
  {Zettl}}]{Kim2013}%
  \BibitemOpen
  \bibfield  {author} {\bibinfo {author} {\bibfnamefont {K.}~\bibnamefont
  {Kim}}, \bibinfo {author} {\bibfnamefont {S.}~\bibnamefont {Coh}}, \bibinfo
  {author} {\bibfnamefont {C.}~\bibnamefont {Kisielowski}}, \bibinfo {author}
  {\bibfnamefont {M.~F.}\ \bibnamefont {Crommie}}, \bibinfo {author}
  {\bibfnamefont {S.~G.}\ \bibnamefont {Louie}}, \bibinfo {author}
  {\bibfnamefont {M.~L.}\ \bibnamefont {Cohen}}, \ and\ \bibinfo {author}
  {\bibfnamefont {A.}~\bibnamefont {Zettl}},\ }\href {\doibase
  10.1038/ncomms3723} {\bibfield  {journal} {\bibinfo  {journal} {Nat.
  Commun.}\ }\textbf {\bibinfo {volume} {4}} (\bibinfo {year} {2013}),\
  10.1038/ncomms3723}\BibitemShut {NoStop}%
\bibitem [{\citenamefont {Bhowmick}\ \emph {et~al.}(2011)\citenamefont
  {Bhowmick}, \citenamefont {Singh},\ and\ \citenamefont
  {Yakobson}}]{Bhowmick2011}%
  \BibitemOpen
  \bibfield  {author} {\bibinfo {author} {\bibfnamefont {S.}~\bibnamefont
  {Bhowmick}}, \bibinfo {author} {\bibfnamefont {A.~K.}\ \bibnamefont {Singh}},
  \ and\ \bibinfo {author} {\bibfnamefont {B.~I.}\ \bibnamefont {Yakobson}},\
  }\href {\doibase 10.1021/jp200671p} {\bibfield  {journal} {\bibinfo
  {journal} {J. Phys. Chem. C}\ }\textbf {\bibinfo {volume} {115}},\ \bibinfo
  {pages} {9889} (\bibinfo {year} {2011})}\BibitemShut {NoStop}%
\bibitem [{\citenamefont {Li}\ and\ \citenamefont {Shenoy}(2011)}]{Li2011_b}%
  \BibitemOpen
  \bibfield  {author} {\bibinfo {author} {\bibfnamefont {J.}~\bibnamefont
  {Li}}\ and\ \bibinfo {author} {\bibfnamefont {V.~B.}\ \bibnamefont
  {Shenoy}},\ }\href {\doibase 10.1063/1.3533804} {\bibfield  {journal}
  {\bibinfo  {journal} {Appl. Phys. Lett.}\ }\textbf {\bibinfo {volume} {98}},\
  \bibinfo {pages} {013105} (\bibinfo {year} {2011})}\BibitemShut {NoStop}%
\bibitem [{\citenamefont {Zhao}\ \emph {et~al.}(2013)\citenamefont {Zhao},
  \citenamefont {Wang}, \citenamefont {Yang}, \citenamefont {Liu},\ and\
  \citenamefont {Liu}}]{Zhao2013}%
  \BibitemOpen
  \bibfield  {author} {\bibinfo {author} {\bibfnamefont {R.}~\bibnamefont
  {Zhao}}, \bibinfo {author} {\bibfnamefont {J.}~\bibnamefont {Wang}}, \bibinfo
  {author} {\bibfnamefont {M.}~\bibnamefont {Yang}}, \bibinfo {author}
  {\bibfnamefont {Z.}~\bibnamefont {Liu}}, \ and\ \bibinfo {author}
  {\bibfnamefont {Z.}~\bibnamefont {Liu}},\ }\href {\doibase
  10.1039/c2cp42994b} {\bibfield  {journal} {\bibinfo  {journal} {Phys. Chem.
  Chem. Phys.}\ }\textbf {\bibinfo {volume} {15}},\ \bibinfo {pages} {803}
  (\bibinfo {year} {2013})}\BibitemShut {NoStop}%
\bibitem [{\citenamefont {Giustino}(2017)}]{FG_review}%
  \BibitemOpen
  \bibfield  {author} {\bibinfo {author} {\bibfnamefont {F.}~\bibnamefont
  {Giustino}},\ }\href {\doibase 10.1103/RevModPhys.89.015003} {\bibfield
  {journal} {\bibinfo  {journal} {Rev. Mod. Phys.}\ }\textbf {\bibinfo {volume}
  {89}},\ \bibinfo {pages} {015003} (\bibinfo {year} {2017})}\BibitemShut
  {NoStop}%
\bibitem [{\citenamefont {Andrea}(2008)}]{Marini2008}%
  \BibitemOpen
  \bibfield  {author} {\bibinfo {author} {\bibfnamefont {M.}~\bibnamefont
  {Andrea}},\ }\href {\doibase 10.1103/PhysRevLett.101.106405} {\bibfield
  {journal} {\bibinfo  {journal} {Phys. Rev. Lett.}\ }\textbf {\bibinfo
  {volume} {101}},\ \bibinfo {pages} {106405} (\bibinfo {year}
  {2008})}\BibitemShut {NoStop}%
\bibitem [{\citenamefont {Giustino}\ \emph {et~al.}(2010)\citenamefont
  {Giustino}, \citenamefont {Louie},\ and\ \citenamefont
  {Cohen}}]{Giustino_2010}%
  \BibitemOpen
  \bibfield  {author} {\bibinfo {author} {\bibfnamefont {F.}~\bibnamefont
  {Giustino}}, \bibinfo {author} {\bibfnamefont {S.~G.}\ \bibnamefont {Louie}},
  \ and\ \bibinfo {author} {\bibfnamefont {M.~L.}\ \bibnamefont {Cohen}},\
  }\href {\doibase 10.1103/PhysRevLett.105.265501} {\bibfield  {journal}
  {\bibinfo  {journal} {Phys. Rev. Lett.}\ }\textbf {\bibinfo {volume} {105}},\
  \bibinfo {pages} {265501} (\bibinfo {year} {2010})}\BibitemShut {NoStop}%
\bibitem [{\citenamefont {Ponc\'e}\ \emph {et~al.}(2014)\citenamefont
  {Ponc\'e}, \citenamefont {Antonius}, \citenamefont {Gillet}, \citenamefont
  {Boulanger}, \citenamefont {Laflamme~Janssen}, \citenamefont {Marini},
  \citenamefont {C\^ot\'e},\ and\ \citenamefont {Gonze}}]{Ponce_2014}%
  \BibitemOpen
  \bibfield  {author} {\bibinfo {author} {\bibfnamefont {S.}~\bibnamefont
  {Ponc\'e}}, \bibinfo {author} {\bibfnamefont {G.}~\bibnamefont {Antonius}},
  \bibinfo {author} {\bibfnamefont {Y.}~\bibnamefont {Gillet}}, \bibinfo
  {author} {\bibfnamefont {P.}~\bibnamefont {Boulanger}}, \bibinfo {author}
  {\bibfnamefont {J.}~\bibnamefont {Laflamme~Janssen}}, \bibinfo {author}
  {\bibfnamefont {A.}~\bibnamefont {Marini}}, \bibinfo {author} {\bibfnamefont
  {M.}~\bibnamefont {C\^ot\'e}}, \ and\ \bibinfo {author} {\bibfnamefont
  {X.}~\bibnamefont {Gonze}},\ }\href {\doibase 10.1103/PhysRevB.90.214304}
  {\bibfield  {journal} {\bibinfo  {journal} {Phys. Rev. B}\ }\textbf {\bibinfo
  {volume} {90}},\ \bibinfo {pages} {214304} (\bibinfo {year}
  {2014})}\BibitemShut {NoStop}%
\bibitem [{\citenamefont {Antonius}\ \emph {et~al.}(2014)\citenamefont
  {Antonius}, \citenamefont {Ponc\'e}, \citenamefont {Boulanger}, \citenamefont
  {C\^ot\'e},\ and\ \citenamefont {Gonze}}]{Antonius_2014}%
  \BibitemOpen
  \bibfield  {author} {\bibinfo {author} {\bibfnamefont {G.}~\bibnamefont
  {Antonius}}, \bibinfo {author} {\bibfnamefont {S.}~\bibnamefont {Ponc\'e}},
  \bibinfo {author} {\bibfnamefont {P.}~\bibnamefont {Boulanger}}, \bibinfo
  {author} {\bibfnamefont {M.}~\bibnamefont {C\^ot\'e}}, \ and\ \bibinfo
  {author} {\bibfnamefont {X.}~\bibnamefont {Gonze}},\ }\href {\doibase
  10.1103/PhysRevLett.112.215501} {\bibfield  {journal} {\bibinfo  {journal}
  {Phys. Rev. Lett.}\ }\textbf {\bibinfo {volume} {112}},\ \bibinfo {pages}
  {215501} (\bibinfo {year} {2014})}\BibitemShut {NoStop}%
\bibitem [{\citenamefont {Ponc{\'e}}\ \emph {et~al.}(2015)\citenamefont
  {Ponc{\'e}}, \citenamefont {Gillet}, \citenamefont {Laflamme~Janssen},
  \citenamefont {Marini}, \citenamefont {Verstraete},\ and\ \citenamefont
  {Gonze}}]{Ponce_2015}%
  \BibitemOpen
  \bibfield  {author} {\bibinfo {author} {\bibfnamefont {S.}~\bibnamefont
  {Ponc{\'e}}}, \bibinfo {author} {\bibfnamefont {Y.}~\bibnamefont {Gillet}},
  \bibinfo {author} {\bibfnamefont {J.}~\bibnamefont {Laflamme~Janssen}},
  \bibinfo {author} {\bibfnamefont {A.}~\bibnamefont {Marini}}, \bibinfo
  {author} {\bibfnamefont {M.}~\bibnamefont {Verstraete}}, \ and\ \bibinfo
  {author} {\bibfnamefont {X.}~\bibnamefont {Gonze}},\ }\href {\doibase
  10.1063/1.4927081} {\bibfield  {journal} {\bibinfo  {journal} {J. Chem.
  Phys.}\ }\textbf {\bibinfo {volume} {143}},\ \bibinfo {pages} {102813}
  (\bibinfo {year} {2015})}\BibitemShut {NoStop}%
\bibitem [{\citenamefont {Lihm}\ and\ \citenamefont {Park}(2020)}]{Lihm_2020}%
  \BibitemOpen
  \bibfield  {author} {\bibinfo {author} {\bibfnamefont {J.-M.}\ \bibnamefont
  {Lihm}}\ and\ \bibinfo {author} {\bibfnamefont {C.-H.}\ \bibnamefont
  {Park}},\ }\href {\doibase 10.1103/PhysRevB.101.121102} {\bibfield  {journal}
  {\bibinfo  {journal} {Phys. Rev. B}\ }\textbf {\bibinfo {volume} {101}},\
  \bibinfo {pages} {121102} (\bibinfo {year} {2020})}\BibitemShut {NoStop}%
\bibitem [{\citenamefont {Miglio}\ \emph {et~al.}(2020)\citenamefont {Miglio},
  \citenamefont {Brousseau-Couture}, \citenamefont {Godbout}, \citenamefont
  {Antonius}, \citenamefont {Chan}, \citenamefont {Louie}, \citenamefont
  {C\^ot\'e}, \citenamefont {Giantomassi},\ and\ \citenamefont
  {Gonze}}]{Gonze_2020}%
  \BibitemOpen
  \bibfield  {author} {\bibinfo {author} {\bibfnamefont {A.}~\bibnamefont
  {Miglio}}, \bibinfo {author} {\bibfnamefont {V.}~\bibnamefont
  {Brousseau-Couture}}, \bibinfo {author} {\bibfnamefont {E.}~\bibnamefont
  {Godbout}}, \bibinfo {author} {\bibfnamefont {G.}~\bibnamefont {Antonius}},
  \bibinfo {author} {\bibfnamefont {Y.-H.}\ \bibnamefont {Chan}}, \bibinfo
  {author} {\bibfnamefont {S.~G.}\ \bibnamefont {Louie}}, \bibinfo {author}
  {\bibfnamefont {M.}~\bibnamefont {C\^ot\'e}}, \bibinfo {author}
  {\bibfnamefont {M.}~\bibnamefont {Giantomassi}}, \ and\ \bibinfo {author}
  {\bibfnamefont {X.}~\bibnamefont {Gonze}},\ }\href {\doibase
  https://doi.org/10.1038/s41524-020-00434-z} {\bibfield  {journal} {\bibinfo
  {journal} {Npj Comput. Mater.}\ }\textbf {\bibinfo {volume} {6}} (\bibinfo
  {year} {2020}),\ https://doi.org/10.1038/s41524-020-00434-z}\BibitemShut
  {NoStop}%
\bibitem [{\citenamefont {Patrick}\ and\ \citenamefont
  {Giustino}(2013)}]{Patrick_2013}%
  \BibitemOpen
  \bibfield  {author} {\bibinfo {author} {\bibfnamefont {C.~E.}\ \bibnamefont
  {Patrick}}\ and\ \bibinfo {author} {\bibfnamefont {F.}~\bibnamefont
  {Giustino}},\ }\href@noop {} {\bibfield  {journal} {\bibinfo  {journal} {Nat.
  Commun.}\ }\textbf {\bibinfo {volume} {4}},\ \bibinfo {pages} {2006}
  (\bibinfo {year} {2013})}\BibitemShut {NoStop}%
\bibitem [{\citenamefont {Patrick}\ and\ \citenamefont
  {Giustino}(2014)}]{Patrick_2014}%
  \BibitemOpen
  \bibfield  {author} {\bibinfo {author} {\bibfnamefont {C.~E.}\ \bibnamefont
  {Patrick}}\ and\ \bibinfo {author} {\bibfnamefont {F.}~\bibnamefont
  {Giustino}},\ }\href@noop {} {\bibfield  {journal} {\bibinfo  {journal} {J.
  Phys.: Condens. Matter}\ }\textbf {\bibinfo {volume} {26}},\ \bibinfo {pages}
  {365503} (\bibinfo {year} {2014})}\BibitemShut {NoStop}%
\bibitem [{\citenamefont {Monserrat}\ and\ \citenamefont
  {Needs}(2014)}]{Bartomeu_2014}%
  \BibitemOpen
  \bibfield  {author} {\bibinfo {author} {\bibfnamefont {B.}~\bibnamefont
  {Monserrat}}\ and\ \bibinfo {author} {\bibfnamefont {R.~J.}\ \bibnamefont
  {Needs}},\ }\href {\doibase 10.1103/PhysRevB.89.214304} {\bibfield  {journal}
  {\bibinfo  {journal} {Phys. Rev. B}\ }\textbf {\bibinfo {volume} {89}},\
  \bibinfo {pages} {214304} (\bibinfo {year} {2014})}\BibitemShut {NoStop}%
\bibitem [{\citenamefont {Zacharias}\ \emph {et~al.}(2015)\citenamefont
  {Zacharias}, \citenamefont {Patrick},\ and\ \citenamefont
  {Giustino}}]{Zacharias_2015}%
  \BibitemOpen
  \bibfield  {author} {\bibinfo {author} {\bibfnamefont {M.}~\bibnamefont
  {Zacharias}}, \bibinfo {author} {\bibfnamefont {C.~E.}\ \bibnamefont
  {Patrick}}, \ and\ \bibinfo {author} {\bibfnamefont {F.}~\bibnamefont
  {Giustino}},\ }\href {\doibase 10.1103/PhysRevLett.115.177401} {\bibfield
  {journal} {\bibinfo  {journal} {Phys. Rev. Lett.}\ }\textbf {\bibinfo
  {volume} {115}},\ \bibinfo {pages} {177401} (\bibinfo {year}
  {2015})}\BibitemShut {NoStop}%
\bibitem [{\citenamefont {Zacharias}\ and\ \citenamefont
  {Giustino}(2016)}]{Zacharias_2016}%
  \BibitemOpen
  \bibfield  {author} {\bibinfo {author} {\bibfnamefont {M.}~\bibnamefont
  {Zacharias}}\ and\ \bibinfo {author} {\bibfnamefont {F.}~\bibnamefont
  {Giustino}},\ }\href {\doibase 10.1103/PhysRevB.94.075125} {\bibfield
  {journal} {\bibinfo  {journal} {Phys. Rev. B}\ }\textbf {\bibinfo {volume}
  {94}},\ \bibinfo {pages} {075125} (\bibinfo {year} {2016})}\BibitemShut
  {NoStop}%
\bibitem [{\citenamefont {Monserrat}(2016)}]{Bartomeu_2016}%
  \BibitemOpen
  \bibfield  {author} {\bibinfo {author} {\bibfnamefont {B.}~\bibnamefont
  {Monserrat}},\ }\href {\doibase 10.1103/PhysRevB.93.100301} {\bibfield
  {journal} {\bibinfo  {journal} {Phys. Rev. B}\ }\textbf {\bibinfo {volume}
  {93}},\ \bibinfo {pages} {100301} (\bibinfo {year} {2016})}\BibitemShut
  {NoStop}%
\bibitem [{\citenamefont {Karsai}\ \emph {et~al.}(2018)\citenamefont {Karsai},
  \citenamefont {Engel}, \citenamefont {Kresse},\ and\ \citenamefont
  {Flage-Larsen}}]{Karsai_2018}%
  \BibitemOpen
  \bibfield  {author} {\bibinfo {author} {\bibfnamefont {F.}~\bibnamefont
  {Karsai}}, \bibinfo {author} {\bibfnamefont {M.}~\bibnamefont {Engel}},
  \bibinfo {author} {\bibfnamefont {G.}~\bibnamefont {Kresse}}, \ and\ \bibinfo
  {author} {\bibfnamefont {E.}~\bibnamefont {Flage-Larsen}},\ }\href@noop {}
  {\bibfield  {journal} {\bibinfo  {journal} {New J. Phys.}\ }\textbf {\bibinfo
  {volume} {20}},\ \bibinfo {pages} {123008} (\bibinfo {year}
  {2018})}\BibitemShut {NoStop}%
\bibitem [{\citenamefont {Zacharias}\ and\ \citenamefont
  {Giustino}(2020)}]{Zacharias_2020}%
  \BibitemOpen
  \bibfield  {author} {\bibinfo {author} {\bibfnamefont {M.}~\bibnamefont
  {Zacharias}}\ and\ \bibinfo {author} {\bibfnamefont {F.}~\bibnamefont
  {Giustino}},\ }\href {\doibase 10.1103/PhysRevResearch.2.013357} {\bibfield
  {journal} {\bibinfo  {journal} {Phys. Rev. Res.}\ }\textbf {\bibinfo {volume}
  {2}},\ \bibinfo {pages} {013357} (\bibinfo {year} {2020})}\BibitemShut
  {NoStop}%
\bibitem [{\citenamefont {Della~Sala}\ \emph {et~al.}(2004)\citenamefont
  {Della~Sala}, \citenamefont {Rousseau}, \citenamefont {G\"orling},\ and\
  \citenamefont {Marx}}]{Della_Sala_2004}%
  \BibitemOpen
  \bibfield  {author} {\bibinfo {author} {\bibfnamefont {F.}~\bibnamefont
  {Della~Sala}}, \bibinfo {author} {\bibfnamefont {R.}~\bibnamefont
  {Rousseau}}, \bibinfo {author} {\bibfnamefont {A.}~\bibnamefont {G\"orling}},
  \ and\ \bibinfo {author} {\bibfnamefont {D.}~\bibnamefont {Marx}},\ }\href
  {\doibase 10.1103/PhysRevLett.92.183401} {\bibfield  {journal} {\bibinfo
  {journal} {Phys. Rev. Lett.}\ }\textbf {\bibinfo {volume} {92}},\ \bibinfo
  {pages} {183401} (\bibinfo {year} {2004})}\BibitemShut {NoStop}%
\bibitem [{\citenamefont {Ram\'{\i}rez}\ \emph {et~al.}(2006)\citenamefont
  {Ram\'{\i}rez}, \citenamefont {Herrero},\ and\ \citenamefont
  {Hern\'andez}}]{Ramirez_2006}%
  \BibitemOpen
  \bibfield  {author} {\bibinfo {author} {\bibfnamefont {R.}~\bibnamefont
  {Ram\'{\i}rez}}, \bibinfo {author} {\bibfnamefont {C.~P.}\ \bibnamefont
  {Herrero}}, \ and\ \bibinfo {author} {\bibfnamefont {E.~R.}\ \bibnamefont
  {Hern\'andez}},\ }\href {\doibase 10.1103/PhysRevB.73.245202} {\bibfield
  {journal} {\bibinfo  {journal} {Phys. Rev. B}\ }\textbf {\bibinfo {volume}
  {73}},\ \bibinfo {pages} {245202} (\bibinfo {year} {2006})}\BibitemShut
  {NoStop}%
\bibitem [{\citenamefont {Ram\'{\i}rez}\ \emph {et~al.}(2008)\citenamefont
  {Ram\'{\i}rez}, \citenamefont {Herrero}, \citenamefont {Hern\'andez},\ and\
  \citenamefont {Cardona}}]{Ramirez_2008}%
  \BibitemOpen
  \bibfield  {author} {\bibinfo {author} {\bibfnamefont {R.}~\bibnamefont
  {Ram\'{\i}rez}}, \bibinfo {author} {\bibfnamefont {C.~P.}\ \bibnamefont
  {Herrero}}, \bibinfo {author} {\bibfnamefont {E.~R.}\ \bibnamefont
  {Hern\'andez}}, \ and\ \bibinfo {author} {\bibfnamefont {M.}~\bibnamefont
  {Cardona}},\ }\href {\doibase 10.1103/PhysRevB.77.045210} {\bibfield
  {journal} {\bibinfo  {journal} {Phys. Rev. B}\ }\textbf {\bibinfo {volume}
  {77}},\ \bibinfo {pages} {045210} (\bibinfo {year} {2008})}\BibitemShut
  {NoStop}%
\bibitem [{\citenamefont {Zacharias}\ \emph {et~al.}(2020)\citenamefont
  {Zacharias}, \citenamefont {Scheffler},\ and\ \citenamefont
  {Carbogno}}]{Zacharias_2020_b}%
  \BibitemOpen
  \bibfield  {author} {\bibinfo {author} {\bibfnamefont {M.}~\bibnamefont
  {Zacharias}}, \bibinfo {author} {\bibfnamefont {M.}~\bibnamefont
  {Scheffler}}, \ and\ \bibinfo {author} {\bibfnamefont {C.}~\bibnamefont
  {Carbogno}},\ }\href {\doibase 10.1103/PhysRevB.102.045126} {\bibfield
  {journal} {\bibinfo  {journal} {Phys. Rev. B}\ }\textbf {\bibinfo {volume}
  {102}},\ \bibinfo {pages} {045126} (\bibinfo {year} {2020})}\BibitemShut
  {NoStop}%
\bibitem [{\citenamefont {Kundu}\ \emph {et~al.}(2021)\citenamefont {Kundu},
  \citenamefont {Govoni}, \citenamefont {Yang}, \citenamefont {Ceriotti},
  \citenamefont {Gygi},\ and\ \citenamefont {Galli}}]{Galli_2021}%
  \BibitemOpen
  \bibfield  {author} {\bibinfo {author} {\bibfnamefont {A.}~\bibnamefont
  {Kundu}}, \bibinfo {author} {\bibfnamefont {M.}~\bibnamefont {Govoni}},
  \bibinfo {author} {\bibfnamefont {H.}~\bibnamefont {Yang}}, \bibinfo {author}
  {\bibfnamefont {M.}~\bibnamefont {Ceriotti}}, \bibinfo {author}
  {\bibfnamefont {F.}~\bibnamefont {Gygi}}, \ and\ \bibinfo {author}
  {\bibfnamefont {G.}~\bibnamefont {Galli}},\ }\href {\doibase
  10.1103/PhysRevMaterials.5.L070801} {\bibfield  {journal} {\bibinfo
  {journal} {Phys. Rev. Materials}\ }\textbf {\bibinfo {volume} {5}},\ \bibinfo
  {pages} {L070801} (\bibinfo {year} {2021})}\BibitemShut {NoStop}%
\bibitem [{\citenamefont {Gorelov}\ \emph {et~al.}(2020)\citenamefont
  {Gorelov}, \citenamefont {Ceperley}, \citenamefont {Holzmann},\ and\
  \citenamefont {Pierleoni}}]{Gorelov2020}%
  \BibitemOpen
  \bibfield  {author} {\bibinfo {author} {\bibfnamefont {V.}~\bibnamefont
  {Gorelov}}, \bibinfo {author} {\bibfnamefont {D.~M.}\ \bibnamefont
  {Ceperley}}, \bibinfo {author} {\bibfnamefont {M.}~\bibnamefont {Holzmann}},
  \ and\ \bibinfo {author} {\bibfnamefont {C.}~\bibnamefont {Pierleoni}},\
  }\href {\doibase 10.1063/5.0031843} {\bibfield  {journal} {\bibinfo
  {journal} {J. Chem. Phys.}\ }\textbf {\bibinfo {volume} {153}},\ \bibinfo
  {pages} {234117} (\bibinfo {year} {2020})}\BibitemShut {NoStop}%
\bibitem [{\citenamefont {Hunt}\ \emph {et~al.}(2020)\citenamefont {Hunt},
  \citenamefont {Monserrat}, \citenamefont {Z\'olyomi},\ and\ \citenamefont
  {Drummond}}]{Hunt_2020}%
  \BibitemOpen
  \bibfield  {author} {\bibinfo {author} {\bibfnamefont {R.~J.}\ \bibnamefont
  {Hunt}}, \bibinfo {author} {\bibfnamefont {B.}~\bibnamefont {Monserrat}},
  \bibinfo {author} {\bibfnamefont {V.}~\bibnamefont {Z\'olyomi}}, \ and\
  \bibinfo {author} {\bibfnamefont {N.~D.}\ \bibnamefont {Drummond}},\ }\href
  {\doibase 10.1103/PhysRevB.101.205115} {\bibfield  {journal} {\bibinfo
  {journal} {Phys. Rev. B}\ }\textbf {\bibinfo {volume} {101}},\ \bibinfo
  {pages} {205115} (\bibinfo {year} {2020})}\BibitemShut {NoStop}%
\bibitem [{\citenamefont {Allen}\ and\ \citenamefont
  {Heine}(1976)}]{Allen_1976}%
  \BibitemOpen
  \bibfield  {author} {\bibinfo {author} {\bibfnamefont {P.~B.}\ \bibnamefont
  {Allen}}\ and\ \bibinfo {author} {\bibfnamefont {V.}~\bibnamefont {Heine}},\
  }\href@noop {} {\bibfield  {journal} {\bibinfo  {journal} {J. Phys. C: Solid
  State Phys.}\ }\textbf {\bibinfo {volume} {9}},\ \bibinfo {pages} {2305}
  (\bibinfo {year} {1976})}\BibitemShut {NoStop}%
\bibitem [{\citenamefont {Allen}\ and\ \citenamefont
  {Cardona}(1981)}]{Allen_Cardona_1981}%
  \BibitemOpen
  \bibfield  {author} {\bibinfo {author} {\bibfnamefont {P.~B.}\ \bibnamefont
  {Allen}}\ and\ \bibinfo {author} {\bibfnamefont {M.}~\bibnamefont
  {Cardona}},\ }\href {\doibase 10.1103/PhysRevB.23.1495} {\bibfield  {journal}
  {\bibinfo  {journal} {Phys. Rev. B}\ }\textbf {\bibinfo {volume} {23}},\
  \bibinfo {pages} {1495} (\bibinfo {year} {1981})}\BibitemShut {NoStop}%
\bibitem [{\citenamefont {Ponc\'e}\ \emph {et~al.}(2016)\citenamefont
  {Ponc\'e}, \citenamefont {Margine}, \citenamefont {Verdi},\ and\
  \citenamefont {Giustino}}]{Ponce_2016_EPW}%
  \BibitemOpen
  \bibfield  {author} {\bibinfo {author} {\bibfnamefont {S.}~\bibnamefont
  {Ponc\'e}}, \bibinfo {author} {\bibfnamefont {E.}~\bibnamefont {Margine}},
  \bibinfo {author} {\bibfnamefont {C.}~\bibnamefont {Verdi}}, \ and\ \bibinfo
  {author} {\bibfnamefont {F.}~\bibnamefont {Giustino}},\ }\href {\doibase
  https://doi.org/10.1016/j.cpc.2016.07.028} {\bibfield  {journal} {\bibinfo
  {journal} {Comput. Phys. Commun.}\ }\textbf {\bibinfo {volume} {209}},\
  \bibinfo {pages} {116 } (\bibinfo {year} {2016})}\BibitemShut {NoStop}%
\bibitem [{\citenamefont {Perdew}\ \emph {et~al.}(1996)\citenamefont {Perdew},
  \citenamefont {Burke},\ and\ \citenamefont {Ernzerhof}}]{GGA_Pedrew_1996}%
  \BibitemOpen
  \bibfield  {author} {\bibinfo {author} {\bibfnamefont {J.~P.}\ \bibnamefont
  {Perdew}}, \bibinfo {author} {\bibfnamefont {K.}~\bibnamefont {Burke}}, \
  and\ \bibinfo {author} {\bibfnamefont {M.}~\bibnamefont {Ernzerhof}},\ }\href
  {\doibase 10.1103/PhysRevLett.77.3865} {\bibfield  {journal} {\bibinfo
  {journal} {Phys. Rev. Lett.}\ }\textbf {\bibinfo {volume} {77}},\ \bibinfo
  {pages} {3865} (\bibinfo {year} {1996})}\BibitemShut {NoStop}%
\bibitem [{\citenamefont {Sun}\ \emph {et~al.}(2015)\citenamefont {Sun},
  \citenamefont {Ruzsinszky},\ and\ \citenamefont {Perdew}}]{SCAN_2015}%
  \BibitemOpen
  \bibfield  {author} {\bibinfo {author} {\bibfnamefont {J.}~\bibnamefont
  {Sun}}, \bibinfo {author} {\bibfnamefont {A.}~\bibnamefont {Ruzsinszky}}, \
  and\ \bibinfo {author} {\bibfnamefont {J.~P.}\ \bibnamefont {Perdew}},\
  }\href {\doibase 10.1103/PhysRevLett.115.036402} {\bibfield  {journal}
  {\bibinfo  {journal} {Phys. Rev. Lett.}\ }\textbf {\bibinfo {volume} {115}},\
  \bibinfo {pages} {036402} (\bibinfo {year} {2015})}\BibitemShut {NoStop}%
\bibitem [{\citenamefont {Giannozzi}\ \emph {et~al.}(2009)\citenamefont
  {Giannozzi} \emph {et~al.}}]{QE}%
  \BibitemOpen
  \bibfield  {author} {\bibinfo {author} {\bibfnamefont {P.}~\bibnamefont
  {Giannozzi}} \emph {et~al.},\ }\href@noop {} {\bibfield  {journal} {\bibinfo
  {journal} {J. Phys.: Condens. Matter}\ }\textbf {\bibinfo {volume} {21}},\
  \bibinfo {pages} {395502} (\bibinfo {year} {2009})}\BibitemShut {NoStop}%
\bibitem [{\citenamefont {Giannozzi}\ \emph {et~al.}(2017)\citenamefont
  {Giannozzi} \emph {et~al.}}]{QE_2}%
  \BibitemOpen
  \bibfield  {author} {\bibinfo {author} {\bibfnamefont {P.}~\bibnamefont
  {Giannozzi}} \emph {et~al.},\ }\href {\doibase 10.1088/1361-648x/aa8f79}
  {\bibfield  {journal} {\bibinfo  {journal} {J. Phys.: Condens. Matter}\
  }\textbf {\bibinfo {volume} {29}},\ \bibinfo {pages} {465901} (\bibinfo
  {year} {2017})}\BibitemShut {NoStop}%
\bibitem [{\citenamefont {Lehtola}\ \emph {et~al.}(2018)\citenamefont
  {Lehtola}, \citenamefont {Steigemann}, \citenamefont {Oliveira},\ and\
  \citenamefont {Marques}}]{LIBXC_2018}%
  \BibitemOpen
  \bibfield  {author} {\bibinfo {author} {\bibfnamefont {S.}~\bibnamefont
  {Lehtola}}, \bibinfo {author} {\bibfnamefont {C.}~\bibnamefont {Steigemann}},
  \bibinfo {author} {\bibfnamefont {M.~J.}\ \bibnamefont {Oliveira}}, \ and\
  \bibinfo {author} {\bibfnamefont {M.~A.}\ \bibnamefont {Marques}},\ }\href
  {\doibase https://doi.org/10.1016/j.softx.2017.11.002} {\bibfield  {journal}
  {\bibinfo  {journal} {SoftwareX}\ }\textbf {\bibinfo {volume} {7}},\ \bibinfo
  {pages} {1 } (\bibinfo {year} {2018})}\BibitemShut {NoStop}%
\bibitem [{\citenamefont {Zacharias}\ and\ \citenamefont
  {Kelires}(2020)}]{Zacharias_2020_3}%
  \BibitemOpen
  \bibfield  {author} {\bibinfo {author} {\bibfnamefont {M.}~\bibnamefont
  {Zacharias}}\ and\ \bibinfo {author} {\bibfnamefont {P.~C.}\ \bibnamefont
  {Kelires}},\ }\href {\doibase 10.1103/PhysRevB.101.245122} {\bibfield
  {journal} {\bibinfo  {journal} {Phys. Rev. B}\ }\textbf {\bibinfo {volume}
  {101}},\ \bibinfo {pages} {245122} (\bibinfo {year} {2020})}\BibitemShut
  {NoStop}%
\bibitem [{\citenamefont {Popescu}\ and\ \citenamefont
  {Zunger}(2012)}]{Popescu_2012}%
  \BibitemOpen
  \bibfield  {author} {\bibinfo {author} {\bibfnamefont {V.}~\bibnamefont
  {Popescu}}\ and\ \bibinfo {author} {\bibfnamefont {A.}~\bibnamefont
  {Zunger}},\ }\href {\doibase 10.1103/PhysRevB.85.085201} {\bibfield
  {journal} {\bibinfo  {journal} {Phys. Rev. B}\ }\textbf {\bibinfo {volume}
  {85}},\ \bibinfo {pages} {085201} (\bibinfo {year} {2012})}\BibitemShut
  {NoStop}%
\bibitem [{\citenamefont {Medeiros}\ \emph {et~al.}(2014)\citenamefont
  {Medeiros}, \citenamefont {Stafstr\"om},\ and\ \citenamefont
  {Bj\"ork}}]{Medeiros_2014}%
  \BibitemOpen
  \bibfield  {author} {\bibinfo {author} {\bibfnamefont {P.~V.~C.}\
  \bibnamefont {Medeiros}}, \bibinfo {author} {\bibfnamefont {S.}~\bibnamefont
  {Stafstr\"om}}, \ and\ \bibinfo {author} {\bibfnamefont {J.}~\bibnamefont
  {Bj\"ork}},\ }\href {\doibase 10.1103/PhysRevB.89.041407} {\bibfield
  {journal} {\bibinfo  {journal} {Phys. Rev. B}\ }\textbf {\bibinfo {volume}
  {89}},\ \bibinfo {pages} {041407} (\bibinfo {year} {2014})}\BibitemShut
  {NoStop}%
\bibitem [{\citenamefont {Freitag}\ \emph {et~al.}(2016)\citenamefont {Freitag}
  \emph {et~al.}}]{Freitag2016}%
  \BibitemOpen
  \bibfield  {author} {\bibinfo {author} {\bibfnamefont {N.~M.}\ \bibnamefont
  {Freitag}} \emph {et~al.},\ }\href {\doibase 10.1021/acs.nanolett.6b02548}
  {\bibfield  {journal} {\bibinfo  {journal} {Nano Lett.}\ }\textbf {\bibinfo
  {volume} {16}},\ \bibinfo {pages} {5798} (\bibinfo {year}
  {2016})}\BibitemShut {NoStop}%
\bibitem [{\citenamefont {Vandersypen}\ \emph {et~al.}(2017)\citenamefont
  {Vandersypen}, \citenamefont {Bluhm}, \citenamefont {Clarke}, \citenamefont
  {Dzurak}, \citenamefont {Ishihara}, \citenamefont {Morello}, \citenamefont
  {Reilly}, \citenamefont {Schreiber},\ and\ \citenamefont
  {Veldhorst}}]{Vandersypen2017}%
  \BibitemOpen
  \bibfield  {author} {\bibinfo {author} {\bibfnamefont {L.~M.~K.}\
  \bibnamefont {Vandersypen}}, \bibinfo {author} {\bibfnamefont
  {H.}~\bibnamefont {Bluhm}}, \bibinfo {author} {\bibfnamefont {J.~S.}\
  \bibnamefont {Clarke}}, \bibinfo {author} {\bibfnamefont {A.~S.}\
  \bibnamefont {Dzurak}}, \bibinfo {author} {\bibfnamefont {R.}~\bibnamefont
  {Ishihara}}, \bibinfo {author} {\bibfnamefont {A.}~\bibnamefont {Morello}},
  \bibinfo {author} {\bibfnamefont {D.~J.}\ \bibnamefont {Reilly}}, \bibinfo
  {author} {\bibfnamefont {L.~R.}\ \bibnamefont {Schreiber}}, \ and\ \bibinfo
  {author} {\bibfnamefont {M.}~\bibnamefont {Veldhorst}},\ }\href
  {https://doi.org/10.1038/s41534-017-0038-y} {\bibfield  {journal} {\bibinfo
  {journal} {Npj Quantum Inf.}\ }\textbf {\bibinfo {volume} {3}},\ \bibinfo
  {pages} {34} (\bibinfo {year} {2017})}\BibitemShut {NoStop}%
\bibitem [{\citenamefont {Quarti}\ \emph {et~al.}(2020)\citenamefont {Quarti},
  \citenamefont {Katan},\ and\ \citenamefont {Even}}]{Quarti2020}%
  \BibitemOpen
  \bibfield  {author} {\bibinfo {author} {\bibfnamefont {C.}~\bibnamefont
  {Quarti}}, \bibinfo {author} {\bibfnamefont {C.}~\bibnamefont {Katan}}, \
  and\ \bibinfo {author} {\bibfnamefont {J.}~\bibnamefont {Even}},\ }\href
  {\doibase 10.1088/2515-7639/aba6f6} {\bibfield  {journal} {\bibinfo
  {journal} {JPhys Mater.}\ }\textbf {\bibinfo {volume} {3}},\ \bibinfo {pages}
  {042001} (\bibinfo {year} {2020})}\BibitemShut {NoStop}%
\bibitem [{\citenamefont {Huang}\ \emph {et~al.}(2021)\citenamefont {Huang},
  \citenamefont {Zacharias}, \citenamefont {Lewis}, \citenamefont {Giustino},\
  and\ \citenamefont {Sharifzadeh}}]{Huang2021}%
  \BibitemOpen
  \bibfield  {author} {\bibinfo {author} {\bibfnamefont {T.~A.}\ \bibnamefont
  {Huang}}, \bibinfo {author} {\bibfnamefont {M.}~\bibnamefont {Zacharias}},
  \bibinfo {author} {\bibfnamefont {D.~K.}\ \bibnamefont {Lewis}}, \bibinfo
  {author} {\bibfnamefont {F.}~\bibnamefont {Giustino}}, \ and\ \bibinfo
  {author} {\bibfnamefont {S.}~\bibnamefont {Sharifzadeh}},\ }\href {\doibase
  10.1021/acs.jpclett.1c00264} {\bibfield  {journal} {\bibinfo  {journal} {J.
  Phys. Chem. Lett.}\ }\textbf {\bibinfo {volume} {12}},\ \bibinfo {pages}
  {3802} (\bibinfo {year} {2021})}\BibitemShut {NoStop}%
\bibitem [{nom()}]{nomad_doi}%
  \BibitemOpen
  \href@noop {} {}\Eprint
  {http://arxiv.org/abs/http://dx.doi.org/10.17172/NOMAD/2021.10.04-1}
  {http://dx.doi.org/10.17172/NOMAD/2021.10.04-1} \BibitemShut {NoStop}%
\end{thebibliography}%


\providecommand{\latin}[1]{#1}
\makeatletter
\providecommand{\doi}
  {\begingroup\let\do\@makeother\dospecials
  \catcode`\{=1 \catcode`\}=2 \doi@aux}
\providecommand{\doi@aux}[1]{\endgroup\texttt{#1}}
\makeatother
\providecommand*\mcitethebibliography{\thebibliography}
\csname @ifundefined\endcsname{endmcitethebibliography}
  {\let\endmcitethebibliography\endthebibliography}{}
\begin{mcitethebibliography}{20}
\providecommand*\natexlab[1]{#1}
\providecommand*\mciteSetBstSublistMode[1]{}
\providecommand*\mciteSetBstMaxWidthForm[2]{}
\providecommand*\mciteBstWouldAddEndPuncttrue
  {\def\EndOfBibitem{\unskip.}}
\providecommand*\mciteBstWouldAddEndPunctfalse
  {\let\EndOfBibitem\relax}
\providecommand*\mciteSetBstMidEndSepPunct[3]{}
\providecommand*\mciteSetBstSublistLabelBeginEnd[3]{}
\providecommand*\EndOfBibitem{}
\mciteSetBstSublistMode{f}
\mciteSetBstMaxWidthForm{subitem}{(\alph{mcitesubitemcount})}
\mciteSetBstSublistLabelBeginEnd
  {\mcitemaxwidthsubitemform\space}
  {\relax}
  {\relax}

\bibitem[Giannozzi \latin{et~al.}(2009)Giannozzi, \latin{et~al.} others]{QE}
Giannozzi,~P., \latin{et~al.}  {QUANTUM} {ESPRESSO}: A Modular and Open-Source
  Software Project for Quantum Simulations of Materials. \emph{J. Phys.:
  Condens. Matter} \textbf{2009}, \emph{21}, 395502\relax
\mciteBstWouldAddEndPuncttrue
\mciteSetBstMidEndSepPunct{\mcitedefaultmidpunct}
{\mcitedefaultendpunct}{\mcitedefaultseppunct}\relax
\EndOfBibitem
\bibitem[Giannozzi \latin{et~al.}(2017)Giannozzi, \latin{et~al.} others]{QE_2}
Giannozzi,~P., \latin{et~al.}  Advanced Capabilities for Materials Modelling
  with Quantum {ESPRESSO}. \emph{J. Phys.: Condens. Matter} \textbf{2017},
  \emph{29}, 465901\relax
\mciteBstWouldAddEndPuncttrue
\mciteSetBstMidEndSepPunct{\mcitedefaultmidpunct}
{\mcitedefaultendpunct}{\mcitedefaultseppunct}\relax
\EndOfBibitem
\bibitem[Bl\"ochl(1994)]{Blochl_1994}
Bl\"ochl,~P.~E. Projector Augmented-Wave Method. \emph{Phys. Rev. B}
  \textbf{1994}, \emph{50}, 17953--17979\relax
\mciteBstWouldAddEndPuncttrue
\mciteSetBstMidEndSepPunct{\mcitedefaultmidpunct}
{\mcitedefaultendpunct}{\mcitedefaultseppunct}\relax
\EndOfBibitem
\bibitem[Kresse and Joubert(1999)Kresse, and Joubert]{Kresse_1995}
Kresse,~G.; Joubert,~D. From Ultrasoft Pseudopotentials to the Projector
  Augmented-Wave Method. \emph{Phys. Rev. B} \textbf{1999}, \emph{59},
  1758--1775\relax
\mciteBstWouldAddEndPuncttrue
\mciteSetBstMidEndSepPunct{\mcitedefaultmidpunct}
{\mcitedefaultendpunct}{\mcitedefaultseppunct}\relax
\EndOfBibitem
\bibitem[Perdew \latin{et~al.}(1996)Perdew, Burke, and
  Ernzerhof]{GGA_Pedrew_1996}
Perdew,~J.~P.; Burke,~K.; Ernzerhof,~M. Generalized Gradient Approximation Made
  Simple. \emph{Phys. Rev. Lett.} \textbf{1996}, \emph{77}, 3865--3868\relax
\mciteBstWouldAddEndPuncttrue
\mciteSetBstMidEndSepPunct{\mcitedefaultmidpunct}
{\mcitedefaultendpunct}{\mcitedefaultseppunct}\relax
\EndOfBibitem
\bibitem[Jianwei \latin{et~al.}(2015)Jianwei, Adrienn, and Perdew]{SCAN_2015}
Jianwei,~S.; Adrienn,~R.; Perdew,~J.~P. Strongly Constrained and Appropriately
  Normed Semilocal Density Functional. \emph{Phys. Rev. Lett.} \textbf{2015},
  \emph{115}, 036402\relax
\mciteBstWouldAddEndPuncttrue
\mciteSetBstMidEndSepPunct{\mcitedefaultmidpunct}
{\mcitedefaultendpunct}{\mcitedefaultseppunct}\relax
\EndOfBibitem
\bibitem[Yao and Kanai(2017)Yao, and Kanai]{SCAN_pseudo}
Yao,~Y.; Kanai,~Y. Plane-Wave Pseudopotential Implementation and Performance of
  SCAN Meta-GGA Exchange-Correlation Functional for Extended Systems. \emph{J.
  Chem. Phys.} \textbf{2017}, \emph{146}, 224105\relax
\mciteBstWouldAddEndPuncttrue
\mciteSetBstMidEndSepPunct{\mcitedefaultmidpunct}
{\mcitedefaultendpunct}{\mcitedefaultseppunct}\relax
\EndOfBibitem
\bibitem[Nocedal and Wright.(2006)Nocedal, and Wright.]{Nocedal_2006}
Nocedal,~J.; Wright.,~S.~J. \emph{Numerical optimization}, 2nd ed.; Springer,
  2006\relax
\mciteBstWouldAddEndPuncttrue
\mciteSetBstMidEndSepPunct{\mcitedefaultmidpunct}
{\mcitedefaultendpunct}{\mcitedefaultseppunct}\relax
\EndOfBibitem
\bibitem[Kunc and Martin(1983)Kunc, and Martin]{Kunc_Martin}
Kunc,~K.; Martin,~R.~M. In \emph{in Ab initio calculation of phonon spectra};
  Devreese,~J., Van~Doren,~V., Van~Camp,~P., Eds.; Plenum: New York, 1983;
  p~65\relax
\mciteBstWouldAddEndPuncttrue
\mciteSetBstMidEndSepPunct{\mcitedefaultmidpunct}
{\mcitedefaultendpunct}{\mcitedefaultseppunct}\relax
\EndOfBibitem
\bibitem[Ackland \latin{et~al.}(1997)Ackland, Warren, and Clark]{Ackland}
Ackland,~G.~J.; Warren,~M.~C.; Clark,~S.~J. \emph{J. Phys.: Condens. Matter}
  \textbf{1997}, \emph{9}, 7861\relax
\mciteBstWouldAddEndPuncttrue
\mciteSetBstMidEndSepPunct{\mcitedefaultmidpunct}
{\mcitedefaultendpunct}{\mcitedefaultseppunct}\relax
\EndOfBibitem
\bibitem[Zacharias and Giustino(2016)Zacharias, and Giustino]{Zacharias_2016}
Zacharias,~M.; Giustino,~F. One-Shot Calculation of Temperature-Dependent
  Optical Spectra and Phonon-Induced Band-Gap Renormalization. \emph{Phys. Rev.
  B} \textbf{2016}, \emph{94}, 075125\relax
\mciteBstWouldAddEndPuncttrue
\mciteSetBstMidEndSepPunct{\mcitedefaultmidpunct}
{\mcitedefaultendpunct}{\mcitedefaultseppunct}\relax
\EndOfBibitem
\bibitem[Sakurai(1994)]{Sakurai}
Sakurai,~J.~J. \emph{Modern Quantum Mechanics}; Addison-Wesley: Reading, 1994;
  pp 285--293\relax
\mciteBstWouldAddEndPuncttrue
\mciteSetBstMidEndSepPunct{\mcitedefaultmidpunct}
{\mcitedefaultendpunct}{\mcitedefaultseppunct}\relax
\EndOfBibitem
\bibitem[Williams(1951)]{Williams_1951}
Williams,~F.~E. Theoretical Low Temperature Spectra of the Thallium Activated
  Potassium Chloride Phosphor. \emph{Phys. Rev.} \textbf{1951}, \emph{82},
  281--282\relax
\mciteBstWouldAddEndPuncttrue
\mciteSetBstMidEndSepPunct{\mcitedefaultmidpunct}
{\mcitedefaultendpunct}{\mcitedefaultseppunct}\relax
\EndOfBibitem
\bibitem[Lax(1952)]{Lax_1952}
Lax,~M. The {F}ranck-{C}ondon Principle and Its Application to Crystals.
  \emph{J. Chem. Phys.} \textbf{1952}, \emph{20}, 1752--1760\relax
\mciteBstWouldAddEndPuncttrue
\mciteSetBstMidEndSepPunct{\mcitedefaultmidpunct}
{\mcitedefaultendpunct}{\mcitedefaultseppunct}\relax
\EndOfBibitem
\bibitem[Zacharias and Giustino(2020)Zacharias, and Giustino]{Zacharias_2020}
Zacharias,~M.; Giustino,~F. Theory of the Special Displacement Method for
  Electronic Structure Calculations at Finite Temperature. \emph{Phys. Rev.
  Res.} \textbf{2020}, \emph{2}, 013357\relax
\mciteBstWouldAddEndPuncttrue
\mciteSetBstMidEndSepPunct{\mcitedefaultmidpunct}
{\mcitedefaultendpunct}{\mcitedefaultseppunct}\relax
\EndOfBibitem
\bibitem[Allen and Heine(1976)Allen, and Heine]{Allen_1976}
Allen,~P.~B.; Heine,~V. Theory of the Temperature Dependence of Electronic Band
  Structures. \emph{J. Phys. C: Solid State Phys.} \textbf{1976}, \emph{9},
  2305--2312\relax
\mciteBstWouldAddEndPuncttrue
\mciteSetBstMidEndSepPunct{\mcitedefaultmidpunct}
{\mcitedefaultendpunct}{\mcitedefaultseppunct}\relax
\EndOfBibitem
\bibitem[Allen and Cardona(1981)Allen, and Cardona]{Allen_Cardona_1981}
Allen,~P.~B.; Cardona,~M. Theory of the Temperature Dependence of the Direct
  Gap of Germanium. \emph{Phys. Rev. B} \textbf{1981}, \emph{23},
  1495--1505\relax
\mciteBstWouldAddEndPuncttrue
\mciteSetBstMidEndSepPunct{\mcitedefaultmidpunct}
{\mcitedefaultendpunct}{\mcitedefaultseppunct}\relax
\EndOfBibitem
\bibitem[Merzbacher(1998)]{Merzbacher}
Merzbacher,~E. \emph{Quantum Mechanics}, 3rd ed.; Johns Wiley \& Sons, Inc.:
  New York, 1998; pp 463--465\relax
\mciteBstWouldAddEndPuncttrue
\mciteSetBstMidEndSepPunct{\mcitedefaultmidpunct}
{\mcitedefaultendpunct}{\mcitedefaultseppunct}\relax
\EndOfBibitem
\bibitem[Ponc{\'e} \latin{et~al.}(2015)Ponc{\'e}, Gillet, Laflamme~Janssen,
  Marini, Verstraete, and Gonze]{Ponce_2015}
Ponc{\'e},~S.; Gillet,~Y.; Laflamme~Janssen,~J.; Marini,~A.; Verstraete,~M.;
  Gonze,~X. Temperature Dependence of the Electronic Structure of
  Semiconductors and Insulators. \emph{J. Chem. Phys.} \textbf{2015},
  \emph{143}, 102813\relax
\mciteBstWouldAddEndPuncttrue
\mciteSetBstMidEndSepPunct{\mcitedefaultmidpunct}
{\mcitedefaultendpunct}{\mcitedefaultseppunct}\relax
\EndOfBibitem
\end{mcitethebibliography}
\end{document}


\maketitle

\newpage 

\section*{Table of Contents}
\begin{itemize}
\item[\textbf{S(I)}] General computational details
\item[\textbf{S(II)}] Derivation of Eq.~(1) using degenerate perturbation theory
\item[\textbf{S(III)}] Evaluation of Eq.~(3) using ZG displacements 
\item[\textbf{S(IV)}] Non-adiabatic effects with ZG displacements 
\end{itemize}
 
\newpage 
\section{\textbf{S(I)} General computational details}
All first-principles calculations are based on density functional theory using 
plane waves basis sets as implemented in {\tt Quantum Espresso}~\cite{QE,QE_2}.
For the majority of our calculations we employed projector augmented wave (PAW) 
pseudopotentials~\cite{Blochl_1994,Kresse_1995} and the Perdew-Burke-Ernzerhof (PBE)~\cite{GGA_Pedrew_1996}
functional for the treatment of electron exchange and correlation.
For comparison purposes we also performed calculations using the strongly constrained 
and appropriately normed (SCAN)~\cite{SCAN_2015} functional together with the 
Troullier-Martins pseudopotential scheme~\cite{SCAN_pseudo}.
The planewaves kinetic energy cutoff was set to 80~Ry. 
The sampling of the Brillouin zone of each graphene quantum dot (GQD) 
was performed using the $\Gamma$ point. To avoid spurious interactions between 
periodic images, a vacuum of more than 20 \AA \, was used along all Cartesian directions
for free-standing (FS) GQDs and perpendicular to the plane of GQDs embedded
in h-BN (GQDs/h-BN). For FS-GQDs we also used the Martyna-Tuckerman approach to limit
any long-ranged Coulomb interactions. All structures were brought to their ground state via 
BFGS optimization~\cite{Nocedal_2006} until the residual force component 
per atom was less than 0.05 eV/\AA. 

The geometries of zigzag-edged hexagonal FS-GQDs with lateral dimensions ($L$) 
up to $2.72$~nm are shown in Figure S1. All dangling bonds of FS-GQDs are passivated with 
hydrogen atoms. For the heterostructures of GQDs/h-BN, we place the GQDs in h-BN matrices so 
that the distance between periodic images of GQDs is larger than 10~\AA. 

To apply the special displacement method (SDM) we perform the following 
steps: (i) the interatomic force constants of each GQD were 
evaluated by means of the frozen-phonon technique~\cite{Kunc_Martin,Ackland} using small 
atomic displacements of $\pm 0.005$~\AA\, along each Cartesian direction, (ii)
the phonon frequencies $\w_\nu$ and eigenmodes $e_{\k\a,\nu}$ were obtained 
by diagonalizing the dynamical matrix, and (iii)
the set of special displacements were obtained via~\cite{Zacharias_2016}:    
\begin{equation}\label{eq.ZG_displ}
   \Delta\tau^{\rm ZG}_{\k\a} = \sqrt{\frac{M_{\rm p}}{M_\k}}\sum_\nu (-1)^{\nu -1} \,  e_{\k\a,\nu} \, \sigma_{\nu,T}, 
\end{equation}
where $\Delta \tau^{\rm ZG}_{\k\a}$ indicates the Zacharias-Giustino (ZG) displacement 
for atom $\k$ along the Cartesian direction $\a$, $M_{\rm p}$ is the proton mass, 
and ${M_\k}$ is the atomic mass. The summation is taken over all phonon modes 
$\nu$ and $\sigma_{\nu,T}$ is the mode-resolved mean square displacement as defined in the manuscript.

\begin{figure}[hbt!]
\includegraphics[width=0.7\textwidth]{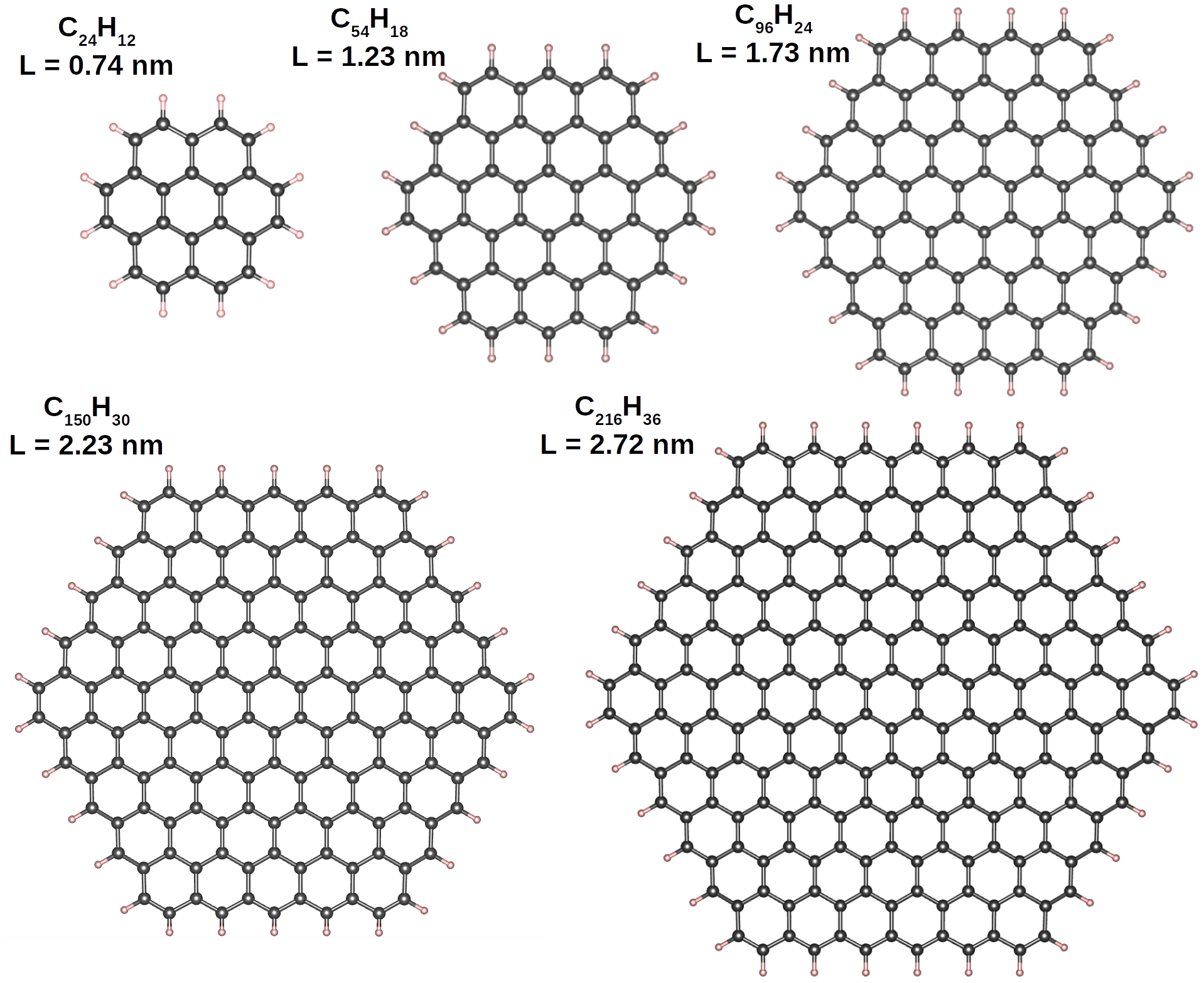}
\begin{flushleft} 
   \noindent  Figure S1:
Ball and stick models of zigzag-edged hexagonal FS-GQDs at their 
relaxed geometries. Dangling bonds of carbon (C) atoms are passivated with hydrogen (H) atoms. 
The lateral dimension and the configuration of each structure are shown. 
\end{flushleft} 
\end{figure}

\newpage 

\section{ \textbf{S(II)} Derivation of Eq.~(1) using degenerate perturbation theory}

Within the non-degenerate Rayleigh-Schr\"odinger perturbation theory~\cite{Sakurai}, an energy level up to its 
second order perturbative expansion is given by: 
\begin{eqnarray} \label{eqnp.1}
\ve'_c &=& \braket{\psi_c|H|\psi_c} + H'_{cc} + \sum_\beta'  \frac{|H'_{\beta c}|^2 }{\ve_c - \ve_\beta}.
\end{eqnarray}
Here, $H$ is the Hamiltonian describing the electronic energy of the system at static equilibrium and 
satisfies $H \ket{\psi_c} = \ve_c \ket{\psi_c}$, where $\ket{\psi_c}$ is the Kohn-Sham wavefunction with energy $\ve_c$. 
The perturbation to the Hamiltonian is represented by $H'$ with matrix elements $H'_{\beta c} = \braket{\psi_\beta|H'|\psi_c}$. 
The primed summation indicates that the term for $\beta=c$ is skipped. 

In the presence of a phonon field, the perturbation to the Hamiltonian up to second order in 
normal coordinates reads:  
\begin{eqnarray} \label{eqnp.2}
H' =  \sum_{\nu} \frac{\partial V_{\rm KS}}{\partial x_\nu} x_\nu 
+ \frac{1}{2}\sum_{\nu \mu} \frac{\partial^2 V_{\rm KS}}{\partial x_\nu \partial x_\mu} x_\nu x_\mu
\end{eqnarray}
where $V_{\rm KS}$ stands for the Kohn-Sham potential. 

Now substituting Eq.~\eqref{eqnp.2} into Eq.~\eqref{eqnp.1} yields: 
\begin{eqnarray} \label{eqnp.3}
\ve'_c = \ve_c + \sum_{\nu} g_{cc\nu} x_\nu + \sum_{\beta \nu \mu}'  \frac{  g_{\beta c\nu} g_{c \beta \mu} }{\ve_c - \ve_\beta} x_\nu x_\mu
+  \frac{1}{2}\sum_{\nu \mu}  \braket{\psi_c | \frac{\partial^2 V_{\rm KS}}{\partial x_\nu \partial x_\mu} | \psi_c } x_\nu x_\mu,
\end{eqnarray}
where we retained terms up to second order in normal coordinates and the electron-phonon matrix element
is defined as $g_{c \beta\nu} = \< \psi_{c} | \D V_{\rm KS}/\D x_\nu | \psi_\beta\> $. 
To obtain the temperature dependence of the energy level renormalization, $\Delta \ve'_c = \ve'_c - \ve_c $, 
we performed the thermal average in the same way as for a Williams-Lax~\cite{Williams_1951,Lax_1952} 
observable. That is~\cite{Zacharias_2020}: 
\begin{eqnarray} \label{eqnp.4}
\Delta \ve_{c, T} = \prod_\nu \int_{-\infty}^{\infty} dx_\nu \frac{\text{exp}(-x^2_\nu/2 \sigma^2_{\nu,T} )}
                     {\sqrt{2\pi}\sigma_{\nu,T}} \Delta \ve'_c.
\end{eqnarray}
Substituting Eq.~\eqref{eqnp.3} into Eq.~\eqref{eqnp.4} and perform the multi-dimensional integration gives: 
\begin{eqnarray}\label{eq.ephcc_nd_AH}
          \DD\ve_{c,T} =  
           {\sum_{\nu \beta}}^\prime \bigg[ \frac{|g_{c \beta\nu}|^2} {\ve_{c}-\ve_\beta} + h_{c \nu}\bigg] 
           \sigma^2_{\nu,T},
\end{eqnarray}
where $h_{c \nu} = 1/2 \, \< \psi_{c} | \D^2 V_{\rm KS}/\partial x^2_\nu | \psi_{c}\>$. 
Equation~\eqref{eq.ephcc_nd_AH} is identical to the second line of Eq.~(1) of the 
main manuscript and represents the renormalization obtained within the non-degenerate adiabatic 
Allen-Heine (AH) theory~\cite{Allen_1976,Allen_Cardona_1981}. We note that the term linear in 
normal coordinates in Eq.~\eqref{eqnp.3} vanishes upon taking the thermal average. 

Now we switch from non-degenerate to degenerate perturbation theory. 
To find the perturbative correction to a two-fold degenerate energy level we need 
to solve the following secular determinant~\cite{Merzbacher}: 
\begin{equation}\label{eqnp.5}
\begin{vmatrix}
    H'_{c_1c_1} - \ve_{c_1}^{(1)} & H'_{c_1c_2} \\ 
    H'_{c_2c_1} & H'_{c_2c_2} - \ve_{c_1}^{(1)}  
\end{vmatrix} = 0.
\end{equation}
Here the subscripts $c_1,c_2$ represent the wavefunctions $\ket{\psi_{c_1}}$ and $\ket{\psi_{c_2}}$, describing the two-fold degenerate level 
with energies $\ve_{c_1} = \ve_{c_2}$, and $\ve_{c_1}^{(1)}$ represents the first order energy correction.  
The solutions of the above equation are:
\begin{eqnarray}\label{eqnp.6}
\ve_{c_1}^{(1)} = \frac{1}{2} \bigg[ \big(H'_{c_1c_1} + H'_{c_2c_2}\big) \pm 
\sqrt{\big(H'_{c_1c_1} + H'_{c_2c_2}\big)^2  - 4 H'_{c_1c_1}H'_{c_2c_2} + 4 H'_{c_1c_2} H'_{c_2c_1}} \bigg]. 
\end{eqnarray}
We now assume that the product of the off-diagonal matrix elements of the perturbation
is negligible next to the product of the diagonal matrix elements, i.e. $H'_{c_1c_2} H'_{c_2c_1} << H'_{c_1c_1}H'_{c_2c_2}$. 
One could also assume, without loss of generality, that $H'$ is diagonal in the degenerate subspace.
Retaining only the linear term in the perturbation of the Hamiltonian [Eq.~\eqref{eqnp.2}],
and performing some straightforward algebra yields:
\begin{eqnarray}\label{eqnp.7}
\ve_{c_1}^{(1)} &=& 
\frac{1}{2} \bigg[ \sum_{\nu} \big(g_{c_1c_1\nu} + g_{c_2c_2 \nu}\big) x_\nu \pm 
\bigg| \sum_{\nu} \big(g_{c_1c_1\nu}  -g_{c_2c_2 \nu}\big) x_\nu \bigg| \bigg].  
\end{eqnarray}
Taking the thermal average as in Eq.~\eqref{eqnp.4} we can readily derive the 
first line of Eq.~(1) in the manuscript. 
At variance with the non-degenerate AH theory, the linear terms
survive the thermal average due to the presence of the absolute value. 

The above analysis can be generalized for an $n$-fold degenerate energy level. 
Following a similar procedure as above, and simplifying the secular determinant 
in a block diagonal form we obtain: 
\begin{eqnarray}\label{eqnp.7}
\Delta \ve_{c_i,T}^{(1)} &=& 
\pm \bigg| \sum_{\nu} \big(g_{c_ic_i\nu}  -g_{c_jc_j \nu}\big) \frac{\sigma_{\nu,T}}{\sqrt{2 \pi}} \bigg|, 
\end{eqnarray}
where $(i,j)$ indicates the pair of states yielding the largest degeneracy 
splitting correction. This result assumes, essentially, that the linear order 
correction for any degenerate level $k \neq i,j$ is zero. 

\newpage

\section{\textbf{S(III)}  Evaluation of Eq.~(3) using ZG displacements}

Within SDM, we apply the set of ZG displacements $\big\{ \Delta\tau^{\rm ZG}_{\k\a}\big\}$ 
using Eq.~\eqref{eq.ZG_displ} to evaluate the energy level renormalization of a cluster as: 
\begin{eqnarray}\label{eq.SI0}
      \DD \ve^{{\rm ZG}}_{c,T } &=&  \sum_\nu 
     \frac{\partial \ve_{c}}{\partial x_\nu}
     \sigma_{\nu,T}  + \frac{1}{2}\sum_\nu \frac{\partial^2 \ve_{c}}{\partial x^2_\nu} \sigma^2_{\nu,T}.
\end{eqnarray}
For the case of degenerate energy levels, ZG displacements lift the degeneracy and the renormalizations of 
the upper ($+$) and lower ($-$) states are given by:
\begin{eqnarray}
      \DD \ve^{{\rm ZG}^+}_{c_1,T } &=&  \sum_\nu 
     \frac{\partial \ve_{c_1}}{\partial x_\nu}
     \sigma_{\nu,T}  + \frac{1}{2}\sum_\nu \frac{\partial^2 \ve_{c_1}}{\partial x^2_\nu} \sigma^2_{\nu,T}
\label{eq.SI1_a}
\\
      \DD \ve^{{\rm ZG}^-}_{c_2,T } &=&  \sum_\nu 
     \frac{\partial \ve_{c_2}}{\partial x_\nu}
     \sigma_{\nu,T}  + \frac{1}{2}\sum_\nu \frac{\partial^2 \ve_{c_2}}{\partial x^2_\nu} \sigma^2_{\nu,T}.
\label{eq.SI1_b}
\end{eqnarray}
The renormalizations of the two states are different due to the different charge densities of 
$\psi_{c_1}$ and $\psi_{c_2}$ (see, for example, Figure S2). The key step for evaluating Eq.~(3)
is to employ the antithetic set of ZG displacements, i.e. $\big\{-\Delta\tau^{\rm ZG}_{\k\a}\big\}$. 
This yields: 
\begin{eqnarray}
      \DD \ve^{{\rm ZG}^-}_{c_1,T } &=&  -\sum_\nu 
     \frac{\partial \ve_{c_1}}{\partial x_\nu}
     \sigma_{\nu,T}  + \frac{1}{2}\sum_\nu \frac{\partial^2 \ve_{c_1}}{\partial x^2_\nu} \sigma^2_{\nu,T}
\label{eq.SI2_a}
\\
      \DD \ve^{{\rm ZG}^+}_{c_2,T } &=&  -\sum_\nu 
     \frac{\partial \ve_{c_2}}{\partial x_\nu}
     \sigma_{\nu,T}  + \frac{1}{2}\sum_\nu \frac{\partial^2 \ve_{c_2}}{\partial x^2_\nu} \sigma^2_{\nu,T},
\label{eq.SI2_b}
\end{eqnarray}
where the energy order of the states is swapped so that $c_1$ corresponds to the lower state, while 
$c_2$ to the upper state. In our calculations, we track the energy order by analysing 
the wavefunctions via projections onto atomic orbitals. 
Combining together Eqs.~\eqref{eq.SI1_a}-\eqref{eq.SI2_b} allows to isolate linear and 
quadratic terms in normal coordinates, i.e:  
\begin{eqnarray}
\frac{\DD \ve^{{\rm ZG}^+}_{c_1,T } + \DD \ve^{{\rm ZG}^-}_{c_1,T }}{2} 
&=& \frac{1}{2}\sum_\nu \frac{\partial^2 \ve_{c_1}}{\partial x^2_\nu} \sigma^2_{\nu,T}
\label{eq.SI3_a}
\\
\frac{\DD \ve^{{\rm ZG}^+}_{c_1,T } - \DD \ve^{{\rm ZG}^-}_{c_1,T }}{2} 
&=& \sum_\nu \frac{\partial \ve_{c_1}}{\partial x_\nu} \sigma_{\nu,T} 
\label{eq.SI3_b}
\\
\frac{ \DD \ve^{{\rm ZG}^-}_{c_2,T } + \DD \ve^{{\rm ZG}^+}_{c_2,T }}{2} 
&=& \frac{1}{2}\sum_\nu \frac{\partial^2 \ve_{c_2}}{\partial x^2_\nu} \sigma^2_{\nu,T}
\label{eq.SI3_c}
\\
\frac{ \DD \ve^{{\rm ZG}^-}_{c_2,T } - \DD \ve^{{\rm ZG}^+}_{c_2,T }}{2} 
&=& \sum_\nu \frac{\partial \ve_{c_2}}{\partial x_\nu} \sigma_{\nu,T}
\label{eq.SI3_d}
\end{eqnarray}
Therefore, to evaluate Eq.~(3) and all terms entering Eq.~(4) of the manuscript we combined 
Eqs.~\eqref{eq.SI3_a}-\eqref{eq.SI3_d} as follows:
\begin{eqnarray}\label{eq.SI4}
\DD \ve^{-}_{c_1,T } &=& -\frac{1}{\sqrt{2 \pi}}\Bigg| \frac{\DD \ve^{{\rm ZG}^+}_{c_1,T } - \DD \ve^{{\rm ZG}^-}_{c_1,T }}{2} 
             - \frac{ \DD \ve^{{\rm ZG}^-}_{c_2,T } - \DD \ve^{{\rm ZG}^+}_{c_2,T }}{2}  \Bigg| 
                     + \frac{\DD \ve^{{\rm ZG}^+}_{c_1,T } + \DD \ve^{{\rm ZG}^-}_{c_1,T }}{2}
\\
\DD \ve^{-}_{c_2,T } &=& -\frac{1}{\sqrt{2 \pi}}\Bigg| \frac{\DD \ve^{{\rm ZG}^+}_{c_1,T } - \DD \ve^{{\rm ZG}^-}_{c_1,T }}{2} 
             - \frac{ \DD \ve^{{\rm ZG}^-}_{c_2,T } - \DD \ve^{{\rm ZG}^+}_{c_2,T }}{2}  \Bigg| 
                     + \frac{\DD \ve^{{\rm ZG}^+}_{c_2,T } + \DD \ve^{{\rm ZG}^-}_{c_2,T }}{2}
\\
\DD \ve^{+}_{v_1,T } &=&\frac{1}{\sqrt{2 \pi}} \Bigg| \frac{\DD \ve^{{\rm ZG}^+}_{v_1,T } - \DD \ve^{{\rm ZG}^-}_{v_1,T }}{2} 
             - \frac{ \DD \ve^{{\rm ZG}^-}_{v_2,T } - \DD \ve^{{\rm ZG}^+}_{v_2,T }}{2}  \Bigg|
                     + \frac{\DD \ve^{{\rm ZG}^+}_{v_1,T } + \DD \ve^{{\rm ZG}^-}_{v_1,T }}{2}
\\
\DD \ve^{+}_{v_2,T } &=&\frac{1}{\sqrt{2 \pi}} \Bigg| \frac{\DD \ve^{{\rm ZG}^+}_{v_1,T } - \DD \ve^{{\rm ZG}^-}_{v_1,T }}{2} 
             - \frac{ \DD \ve^{{\rm ZG}^-}_{v_2,T } - \DD \ve^{{\rm ZG}^+}_{v_2,T }}{2}  \Bigg|
                     + \frac{\DD \ve^{{\rm ZG}^+}_{v_2,T } + \DD \ve^{{\rm ZG}^-}_{v_2,T }}{2}.
\end{eqnarray}

To minimize the numerical error in our SDM calculations we performed configurational averaging 
as described in ref~\cite{Zacharias_2016} . Our calculated zero-point renormalization 
of GQD/h-BN ($L=2.23$~nm) using up to 32 ZG antithetic pairs is shown in Figure S3. Our results 
show that the gap renormalization is converged within 10~meV.

\begin{figure}[htb!]
\includegraphics[width=0.95\textwidth]{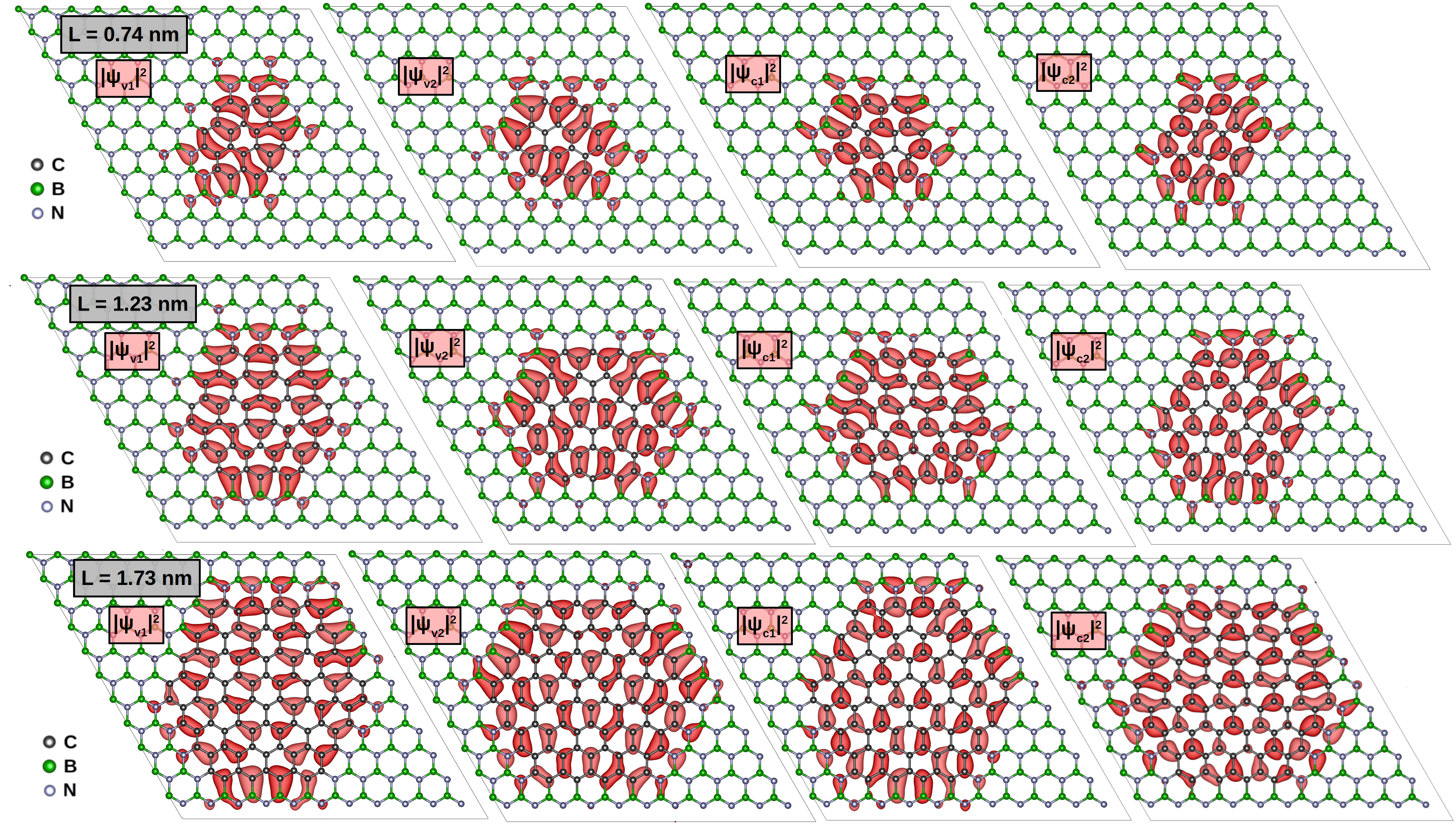}
\begin{flushleft} 
   \noindent  Figure S2:
Charge density of the twofold degenerate Kohn-Sham states of the valence band maximum 
($v_1, v_2$) and conduction band minimum ($c_1, c_2$) of GQDs/h-BN. 
The isovalue is set to 1\% of the maximum. In each case we indicate the size $L$ of the dot. 
\end{flushleft} 
\end{figure}

\begin{figure}[htb!]
\includegraphics[width=0.55\textwidth]{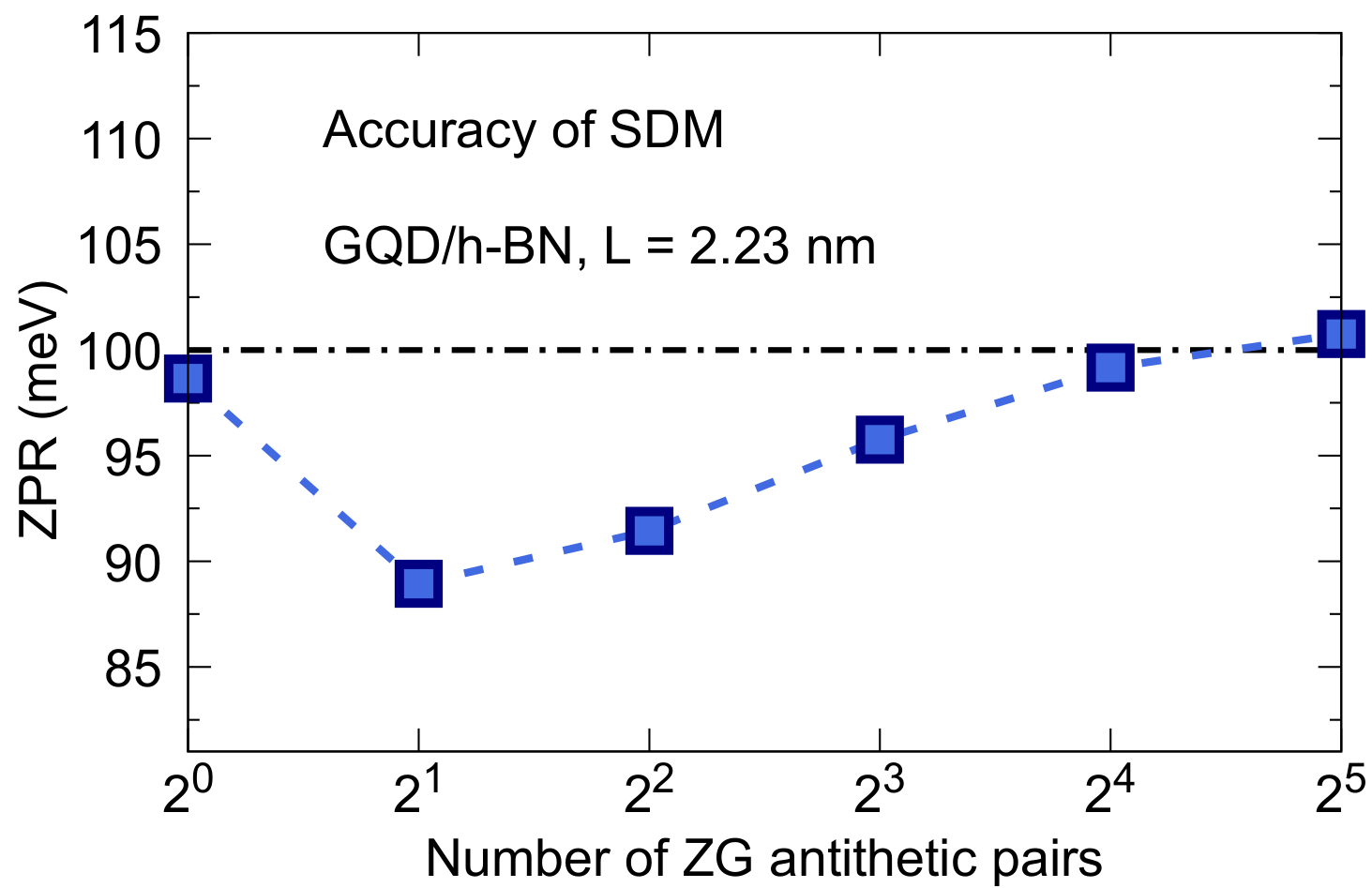}
\begin{flushleft} 
   \noindent  Figure S3:
Variation of the zero-point renormalization (ZPR) of the GQD/h-BN ($L=2.23$~nm) 
with the number of ZG antithetic pairs. The horizontal dashed line indicates the 
converged value of the ZPR.
\end{flushleft} 
\end{figure}

\newpage 
\, 
\newpage
\section{\textbf{S(IV)}  Non-adiabatic effects with ZG displacements}
Following ref~\cite{Ponce_2015} the expression for the non-adiabatic (and non-degenerate) AH energy level 
renormalization reads: 
\begin{equation}\label{eq.ephcc_AH_nonad}
   \DD\ve^{\rm na}_{c} =  {\sum_{\nu \beta}}^\prime \bigg[ \frac{|g_{c \beta\nu}|^2} 
{\ve_{c}-\ve_\beta - \hbar \omega_\nu} + h_{c \nu}\bigg] \sigma^2_{\nu,T},
\end{equation}
where ${\rm na}$ stands for non-adiabatic. Neglecting the phonon-energies in the denominator recovers 
the adiabatic AH theory; this is justified for systems whose band gap energy is much larger than 
their phonon frequencies. 

Now, for non-dispersive band edges (as for GQDs; see Figure~3 of main manuscript) we can replace
the energy denominators with $\ve_g = \ve_c - \ve_v$ and thus write for the band gap renormalization:
\begin{equation}\label{eq.ephcc_AH_nonad_1}
   \DD \ve^{\rm na}_{g} =  {\sum_{\nu}} \bigg[ 2 \frac{|g_{c v \nu}|^2} 
{\ve_g - \hbar \omega_\nu} + h_{c \nu} - h_{v \nu}\bigg] \sigma^2_{\nu,T}.
\end{equation}
If we proceed to a further simplification that $h_{c \nu} - h_{v \nu}$ is negligible 
next to $2|g_{c v \nu}|^2/(\ve_g - \hbar \omega_\nu)$, then we can write: 
\begin{equation}\label{eq.ephcc_AH_nonad_3}
   \DD\ve^{\rm na}_{g} =  {\sum_{\nu}} \bigg[ 2 \frac{|g_{c v \nu}|^2} 
{\ve_g - \hbar \omega_\nu} \bigg] \sigma^2_{\nu,T}.
\end{equation}
Using the same simplifications, the adiabatic (ad) AH expression reads: 
\begin{equation}\label{eq.ephcc_AH_nonad_4}
   \DD\ve^{\rm ad}_{g} =  {\sum_{\nu}} \bigg[ 2 \frac{|g_{c v \nu}|^2} 
{\ve_g} \bigg] \sigma^2_{\nu,T}.
\end{equation}
Comparing Eqs.~\eqref{eq.ephcc_AH_nonad_3} and~\eqref{eq.ephcc_AH_nonad_4}, non-adiabatic effects 
can be incorporated in ZG displacements by modifying the mode-resolved mean square displacements as: 
\begin{equation}\label{eq.ephcc_AH_nonad_5}
\sigma^{\rm na}_{\nu,T} = \sqrt{\frac{\ve_g} { (\ve_g - \hbar \omega_\nu)}} \, \sigma_{\nu,T}. 
\end{equation}
We note that non-adiabaticity does not affect the degeneracy splitting term [first line of Eq.~(1) in
the manuscript]. Hence, for evaluating the energy level renormalization with non-adiabatic effects, 
we combine (i) the adiabatic versions of Eqs~\eqref{eq.SI3_b} and \eqref{eq.SI3_d} with (ii)
the non-adiabatic versions of Eqs~\eqref{eq.SI3_a} and \eqref{eq.SI3_c}.

\bibliography{references}{}